\documentclass[aps,prb,twocolumn,showpacs,groupedaddress,letterpaper]{revtex4}
\usepackage{graphicx}
\usepackage{amssymb}
\usepackage{amsmath}
\newlength{\figwidth}
\usepackage{times}
\begin{document}
\preprint{\today}
\title{Universal Scaling of the Quantum Conductance \\ 
of an Inversion-Symmetric Interacting Model}

\author{Axel Freyn}
\affiliation{ Institut N\'eel, 25 avenue des Martyrs, BP 166, 38042 Grenoble, France}
\affiliation{Service de Physique de l'\'Etat Condens\'e (CNRS URA 2464), 
IRAMIS/SPEC, CEA Saclay, 91191 Gif-sur-Yvette, France}
 
\author{Jean-Louis Pichard}
\affiliation{Service de Physique de l'\'Etat Condens\'e (CNRS URA 2464), 
IRAMIS/SPEC, CEA Saclay, 91191 Gif-sur-Yvette, France}

\begin{abstract} 

We consider quantum transport of spinless fermions in a 1D lattice 
embedding an interacting region (two sites with inter-site repulsion 
$U$ and inter-site hopping $t_d$, coupled to leads by hopping terms $t_c$). 
Using the numerical renormalization group for the particle-hole symmetric 
case, we study the quantum conductance $g$ as a function of the inter-site 
hopping $t_d$. The interacting region, which is perfectly reflecting when 
$t_d \to 0$ or $t_d \to \infty$, becomes perfectly transmitting if $t_d$ 
takes an intermediate value $\tau(U,t_c)$ which defines the characteristic 
energy of this interacting model. When $t_d < t_c \sqrt{U}$, $g$ is given by 
a universal function of the dimensionless ratio $X=t_d/\tau$. This universality 
characterizes the non-interacting regime where $\tau=t_c^2$, the perturbative 
regime ($U < t_c^2$) where $\tau$ can be obtained using Hartree-Fock theory, and 
the non-perturbative regime ($U > t_c^2$) where $\tau$ is twice the characteristic 
temperature $T_K$ of an orbital Kondo effect induced by the inversion symmetry. 
When $t_d < \tau$, the expression $g(X)=4(X+X^{-1})^{-2}$ valid without 
interaction describes also the conductance in the presence of the interaction. 
To obtain those results, we map this spinless model onto an Anderson model with 
spins, where the quantum impurity is at the end point of a semi-infinite 1D lead 
and where $t_d$ plays the role of a magnetic field $h$. This allows us to describe 
$g(t_d)$ using exact results obtained for the magnetization $m(h)$ of the Anderson 
model at zero temperature. We expect this universal scaling to be valid also in models 
with 2D leads, and observable using 2D semi-conductor heterostructures and an 
interacting region made of two identical quantum dots with strong capacitive 
inter-dot coupling and connected via a tunable quantum point contact.

\end{abstract}
\pacs{71.10.-w,72.10.-d,73.23.-b} 

\maketitle
\section*{Introduction}
 In quantum transport theory, a nanosystem inside which the electrons do not 
interact has a zero temperature conductance which is given (in units of the 
conductance quantum $e^2/h$ for spin polarized electrons, $2e^2/h$ with spin 
degeneracy) by 
\begin{equation}
g= |t_{ns}(E_F)|^2, 
\end{equation}
where $|t_{ns}(E_F)|^2$ is the probability for an electron at the Fermi energy 
$E_F$ to be transmitted through the nanosystem. This Landauer-B\"uttiker formula can 
be extended \cite{meir-wingreen,molina1} to an interacting nanosystem, 
if it behaves as a non-interacting nanosystem with renormalized parameters as the 
temperature $T \to 0$. However, this effective non-interacting nanosystem does not 
describe only the interacting region, but depends also on the presence of scatterers 
which can be outside, in the attached leads. This non-local aspect of the effective 
transmission $|t_{\it ns}(E_F)|^2$ is characteristic of nanosystems inside which 
electrons interact and has been studied in 1D models \cite{molina2,weinmann} 
using the density matrix renormalization group (DMRG), and in 1D \cite{asada,freyn-pichard1,freyn-pichard2} 
and 2D models \cite{freyn-pichard3} using the Hartree-Fock (HF) approximation.

In this work, we study how the effective transmission of a nanosystem with perfect leads is  
renormalized by local interactions acting inside the nanosystem, using the numerical 
renormalization group (NRG) algorithm \cite{krishna-murthy1,krishna-murthy2,krishna-murthy3,
bulla,hewson} and an inversion-symmetric interacting model (ISIM). This model describes the 
scattering of spin-polarized electrons (spinless fermions) by an interacting region (two 
sites characterized by an inter-site hopping term $t_d$, coupling terms $t_c$ and an 
inter-site repulsion $U$). Our study is restricted to the symmetric case 
(i.e.\ the case where ISIM is invariant under particle-hole symmetry). 

Firstly, we prove that ISIM, which is perfectly reflecting when the inter-site hopping term 
$t_d \to 0$ or $t_d \to \infty$, exhibits a peak of perfect transmission for an intermediate 
value $\tau(U,t_c)$ of $t_d$. This scale $\tau(U,t_c)$ defines very precisely 
the fundamental energy scale of ISIM. HF theory gives correctly this peak of 
perfect transmission when $U < t_c^2$, but does not give it when $U$ exceeds 
$t_c^2$, showing the existence of a non-perturbative regime where 
the use of the NRG algorithm is required. In this non-perturbative regime,
\begin{equation}
\tau(U,t_c)= 2 T_K\,,
\label{Eq-tau}
\end{equation}
where $T_K$ is the characteristic temperature of an orbital Kondo effect 
induced by the inversion symmetry. 

Secondly, we show that the zero-temperature conductance $g$ is given by a universal 
function $g(X)$ of the dimensionless coordinate $X=t_d/\tau$. This function $g(X)$ is 
independent of the choice of $t_c$ and $U$ as far as $t_d <t_c \sqrt{U}$. When $t_d < \tau$, 
\begin{equation}
g(X)=4 (X+X^{-1})^{-2}\,.
\end{equation}
When $\tau < t_d < t_c \sqrt{U}$, the conductance $g(X)$ can be described  
by another function 
\begin{equation}
g(X) \approx \sin^2\left( \pi\left(2.02-\frac{0.74}{\ln (2.8 X)}+ \ldots\right )\right)\,, 
\label{g-lm-1}
\end{equation}
which is related to an exact result obtained by Tsvelick and Wiegmann
\cite{tsvelick-wiegmann1,tsvelick-wiegmann2} with Bethe-Ansatz for the 
magnetization of the Anderson model at zero temperature. When $t_d$ exceeds $t_c\sqrt{U}$, 
the interaction becomes irrelevant, $\tau$ is given by its non-interacting value $t_c^2$, 
$X=t_d/t_c^2$ and $g(X)=4(X+X^{-1})^{-2}$ again. The conductance $g$ is one example of the 
physical properties of ISIM which are given by universal functions of $t_d/\tau$ at zero 
temperature. The low-energy effective one-body excitations provide another example for 
which we show the corresponding universal curves.    

In order to obtain these universal functions, it is useful to notice that the 
inversion symmetry of ISIM gives rise to a pseudo-spin, allowing to exactly map 
this 1D spinless model onto an Anderson model with spins where the inter-site 
hopping $t_d$ plays the role of a magnetic field, the quantum impurity being 
at the end point of a single semi-infinite chain. Therefore, the behavior of ISIM as 
a function of $t_d$ is related to the behavior of the Anderson model as a function of 
an applied magnetic field $h$. 

 The paper is organized as follows: Section~\ref{Anderson-Kondo}  introduces universal 
aspects which characterize the Anderson model and are relevant for quantum dots where spin or 
orbital Kondo effects occur. The studied spinless model with inversion symmetry (ISIM) 
is defined in section~\ref{section-ISIM} and mapped onto an Anderson model with magnetic 
field in section~\ref{Equivalent Anderson model}. A second transformation is performed 
in section~\ref{NRG Chain}, based on the usual logarithmic discretization of the energy 
band of the leads, to get the final model used for the NRG study. The study is restricted 
to the case with particle-hole symmetry in section~\ref{symmetric case}. The low energy 
excitations are first considered as a function of the temperature in section~\ref{Three fixed 
points}. When $U > \pi \Gamma$, $\Gamma=t_c^2$ being the level width of the scattering region, 
the 3 fixed points [free orbital (FO), local moment (LM) and strong coupling (SC)] characterizing 
the Anderson model without field are recovered when $t_d$ is small enough. As $t_d$ increases, 
the LM fixed point characterizing the model above the Kondo temperature $T_K$ disappears. In 
section~\ref{free fermion fixed points}, we study the low energy excitations of ISIM as a function 
of $t_d$ in the limit $T \to 0$. We find that they can always be described by a set of effective 
one-body excitations, showing that a continuous line of free fermion fixed points goes from the 
SC limit of the Anderson model ($t_d=0$) towards a new simple limit: the polarized orbital (PO) 
fixed point ($t_d \to \infty$). Between the SC and PO fixed points, we show in 
section~\ref{Characteristic Energy Scale} that there is always an intermediate value $\tau(U,t_c)$ of $t_d$ 
for which ISIM is perfectly transmitting. $\tau(U,t_c)$ defines the fundamental energy scale of ISIM. 
In section~\ref{quantum conductance}, a simple method for calculating $g(t_d)$ from the effective one 
body excitations characterizing ISIM when $T \to 0$ is introduced. Using this method, we give in 
section~\ref{quantum conductance} the main result of this work, i.e., if one uses the dimensionless ratio 
$X=t_d/\tau$, the physical properties (conductance or effective one-body spectra) are universal and 
independent of $t_c$ and $U$ as far as $t_d < t_c \sqrt{U}$. This universal regime is divided in a 
first regime where the system is not very far from the SC fixed point ($t_d \leq \tau$, 
subsection~\ref {SC limit}) and where $g(X)=4(X+X^{-1})^{-2}$, followed by a second regime where $g(X)$ 
is given by another universal function ($\tau \leq t_d < t_c\sqrt{U}$, subsection~\ref{LM limit}). 
In the equivalent Anderson model, this second regime is characterized by the occurrence of a magnetic 
moment. When $t_d > t_c\sqrt{U}$, the interaction $U$ becomes irrelevant and $g(X)=4(X+X^{-1})^{-2}$ with 
$X=t_d/t_c^2$ (subsection~\ref{PO limit}). In section~\ref{Perturbative regime}, we show that 
HF theory gives the values of $g$ obtained from the NRG spectra and the scale $\tau(U,t_c)$, if $U$ 
does not exceed $t_c^2$. In contrast, HF theory fails to give perfect transmission if $U > t_c^2$, showing 
the existence of a non-perturbative regime for ISIM. To obtain $\tau$ in the non-perturbative regime, we 
first revisit in section~\ref{Friedel Sum Rule} a method giving $g$ from the difference of occupation numbers 
between the even and odd orbitals of the nanosystem. This method based on Friedel sum rule (FSR) contains an 
assumption. If ISIM is near the SC fixed point ($t_d < \tau$, non-perturbative regime), this assumption turns 
out to be justified. In that case, $g(t_d)$ can be obtained from the impurity magnetization $m(h)$ of the 
equivalent Anderson model with a magnetic field $h$ at zero temperature. Exact results giving $m(h)$ are 
reviewed in section~\ref{Bethe-Ansatz} for the Anderson model. Using those results, we show in 
section~\ref{non-perturbative} that $\tau(U,t_c)=2 T_K$ in the non-perturbative regime, $T_K$ being the 
characteristic temperature of the orbital Kondo effect exhibited by ISIM when $U > t_c^2$. Moreover, a fit 
inspired from the exact behavior of $m(h)$ in the local moment regime is used for describing the universal 
function $g(X)$ when $\tau \leq t_d <t_c\sqrt{U}$. Eventually, we summarize in section~\ref{Conclusion} the 
universal aspects obtained using a simple 1D model, and we conjecture that they can be extended to 2D 
models and observed in 2D semi-conductor heterostructures, where the nanosystem would consist of two 
identical quantum dots coupled by a quantum point contact.  

\section{Anderson model, Kondo Physics, Quantum Dots and Universality}
\label{Anderson-Kondo} 

 The Anderson model describes a single site with Hubbard interaction $U$ coupled to a 3D bath of 
conduction electrons. This is one of the quantum impurity models \cite{hewson} which were introduced 
to study the resistance minimum observed in metals with magnetic impurities. The Kondo problem refers 
to the failure of perturbative techniques to describe this minimum. The solution of these models by 
the NRG algorithm, a non-perturbative technique \cite{krishna-murthy1,krishna-murthy2,krishna-murthy3,
bulla,hewson} introduced by Wilson, is at the origin of the discovery of universal behaviors which can 
emerge from many-body effects. Without magnetic field $h$ and with particle-hole symmetry \cite{krishna-murthy2}, 
the Anderson model maps onto the Kondo Hamiltonian if $U>\pi \Gamma$, $\Gamma\propto t_c^2$ being the 
impurity-level width. In that case, there is a non-perturbative regime where the temperature dependence of 
physical observables such as the impurity susceptibility is given by universal functions of $T/T_K$, 
$T_K$ being the Kondo temperature. If $U < \pi \Gamma$, the impurity susceptibility can be obtained by 
perturbation theory. Universality characterizes not only the behavior of the Anderson model as a function 
of the temperature $T$, but also its behavior at $T=0$ as a function of an applied magnetic field $h$. 
Using the Bethe-Ansatz, Tsvelick and Wiegmann \cite{tsvelick-wiegmann1,tsvelick-wiegmann2} have obtained 
for the magnetization $m(h)$ a universal function of the dimensionless variable $h/T_K$ when $T \to 0$. 

The possibility to design artificial magnetic impurities in nanoscale conductors has opened 
\cite{goldhaber-gordon,cronenwett} a second era for quantum impurity models. Measuring the 
conductance $g$ of quantum dots created by electrostatic gates, in a 2D electron gas 
\cite{goldhaber-gordon,cronenwett,grobis} or in carbon nanotubes \cite{delattre}, one obtains values 
which can be on universal curves as functions of $T/T_K$ if there is a Kondo effect. 
Moreover, quantum dots open the possibility to study the Kondo effect as a function of the coupling between 
the impurity and the continuum of conduction electrons, and not only as functions of the temperature and 
of the magnetic field. As pointed out in Ref.~\onlinecite{kaul}, this gives the opportunity to do the spectroscopy 
of the Kondo problem. Notably, the weak to strong coupling crossover can be studied by varying gate voltages, 
when metallic gates are used for creating quantum dots. Kondo physics was first related to the antiferromagnetic 
coupling between a magnetic impurity and the spin of the host's conduction electrons. This is why Kondo physics was 
first expected and seen \cite{goldhaber-gordon,cronenwett} in quantum dots with odd numbers of electrons, weakly 
coupled to leads. However, it was realized that a localized electronic state coupled to a continuum can give rise 
to a large class of different Kondo effects, including the original spin-$1/2$ Kondo effect, various orbital Kondo 
effects and the SU(4) Kondo effect occurring if a spin Kondo effect co-exists with an orbital Kondo effect. 
 
In this framework, ISIM is a model which can be used for describing the quantum conductance of spin polarized 
electrons in an inversion-symmetric double-dot setup with strong capacitive inter-dot coupling, as a function of 
the inter-dot hopping $t_d$. For such a setup, $t_d$ could be varied by electrostatic gates if the two dots are 
coupled by a quantum point contact, and our study describes the effect of this coupling upon the orbital Kondo 
effect induced by the inversion symmetry. Eventually, universal aspects of many-body phenomena characterize 
not only equilibrium quantum transport, but also non-equilibrium quantum transport which occurs in the presence of 
a large source-drain bias $V_{sd}$. Measures of the conductance \cite{grobis} and of the current noise \cite{delattre} 
of Kondo dots have recently confirmed the expected universality if one measures $T$ or $V_{sd}$ in units of $T_K$. 
We describe here another universal aspect of linear quantum transport, i.e., the quantum conductance of a spin polarized 
inversion symmetric double-dot setup should be a universal function of the dimensionless inter-dot hopping $t_d/T_K$ 
when $T \to 0$. 

Kondo physics is also at the origin of spinless models, as the interacting resonant level model \cite{mehta} (IRLM) 
which describes a resonant level ($V_d d^{\dagger}d$) coupled to two baths of spinless electrons via tunneling 
junctions and an interaction $U$ between the level and the baths. IRLM, which is now used for studying non-equilibrium 
quantum transport \cite{mehta,boulat}, is related to the Kondo model, the charge states $n_d=0,1$ playing the role 
of spin states. Both ISIM and IRLM are inversion symmetric. However, the Zeeman field acting on the impurity is played 
by the hopping term $t_d$ for ISIM, and by the site energy $V_d$ for IRLM. Therefore, ISIM does not transmit the electrons 
without ``field'', while IRLM does. Though we study in this work a finite density of particles, let us mention that the 
two-particle scattering problem has been solved \cite{dhar} for ISIM. 
 
\section{Inversion-Symmetric Interacting Model}
\label{section-ISIM} 

\begin{figure}
\centerline{
\includegraphics[width=\columnwidth]{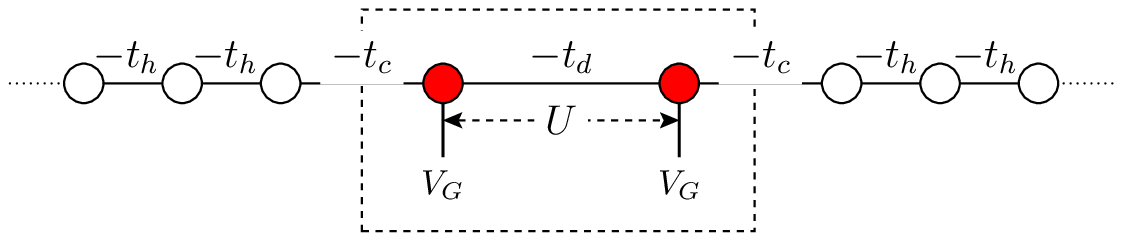}
}
\caption
{(Color online) Studied setup (ISIM) where spin polarized electrons 
(spinless fermions) can be scattered by a nanosystem made of the 
2 red sites (energy $V_G$, inter-site repulsion $U$ and 
an internal hopping $t_d$). The nanosystem is embedded by 
coupling terms $t_c$ into a 1D lattice (hopping term $t_h$). 
}  
\label{FIGURE1} 
\end{figure}

The ISIM model is sketched in Fig.~\ref{FIGURE1} and consists of a 1D 
tight binding lattice (hopping term $t_h$) where a finite density of spin polarized 
electrons (spinless fermions) can be scattered by a central region made of 2 sites 
of potential $V_G$, with an internal hopping term $t_d$, and 2 coupling terms $t_c$. 
The difficulty comes from the presence of a repulsion of strength $U$ which acts 
if the two sites of the central region are occupied. 

The ISIM Hamiltonian reads:
\begin{equation} 
H=H_{ns}+H_{c}+H_{l}\,,
\end{equation}
where the Hamiltonian of the central region (the interacting nanosystem) is given by
\begin{equation}
\label{eq_ham_S}
H_{ns} = - t_d \left( c_0^\dagger c_1^{\vphantom{\dagger}} + c_1^\dagger
c_0^{\vphantom{\dagger}} \right ) 
+ V_G \left( n_0 + n_1 \right ) + U n_0 n_1\,.
\end{equation}
$c_x^\dagger$ and $c_x^{\vphantom{\dagger}}$ are spinless fermion operators 
at site $x$ and $n_x=c_x^\dagger c_x^{\vphantom{\dagger}}$. The coupling 
Hamiltonian between the nanosystem and the leads reads
\begin{equation}
H_{c}=- t_c ( c_{-1}^\dagger c_0^{\vphantom{\dagger}} + c_1^\dagger 
c_2^{\vphantom{\dagger}} + H.c. )\,, 
\end{equation}
while the leads are described by an Hamiltonian 
\begin{equation}
H_{l}= - t_h 
\sideset{}{'}{\sum}_{x=-\infty}^{\infty} ( c_x^\dagger 
c_{x+1}^{\vphantom{\dagger}} + H.c.)\,, 
\end{equation}
where $\sideset{}{'}{\sum}$ means 
that $x=-1,0,1$ are omitted from the summation. 

\begin{figure}
\centerline{
\includegraphics[width=0.8\columnwidth]{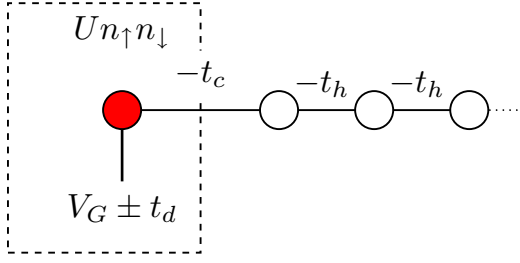}
}
\caption
{(Color online) Equivalent Anderson model: Electrons with a pseudo spin ($\uparrow = 
\text{even}$, $\downarrow = \text{odd}$) are free to move on a semi-infinite chain 
(hopping term $t_h$) with a quantum impurity (red site) attached (hopping term $t_c$) 
to its end point. The inter-site ISIM interaction becomes a Hubbard 
interaction $U n_{\uparrow} n_{\downarrow}$ between impurity orbitals 
of different pseudo-spins. The impurity potential $V_G$ has now a Zeeman 
term $\pm t_d$.
}  
\label{FIGURE2} 
\end{figure}
\begin{figure}
\centerline{
\includegraphics[width=0.8\columnwidth]{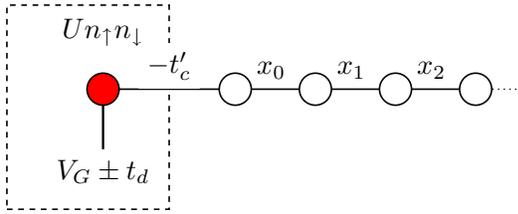}
}
\caption
{(Color online) Corresponding NRG chain: The quantum impurity (red site) 
is now coupled via an hopping term $t'_c$ (Eq.~\eqref{coupling_NRG}) to a 
1D lattice, where the sites are labelled by $n$ and describe conduction electron 
excitations of length scale $\Lambda^{n/2} k_F^{-1}$ centered on the 
impurity. The successive sites are now coupled via hopping terms 
(Eq.~\eqref{hopping_NRG}) which fall off as $\Lambda^{-n/2}$. Used 
discretization parameter $\Lambda=2$.
}  
\label{FIGURE3} 
\end{figure}

\section{Equivalent Anderson model with magnetic field} 
\label{Equivalent Anderson model}

Because of inversion symmetry, one can map ISIM onto a single semi-infinite 1D 
lattice where the fermions have a pseudo-spin and where the double-site nanosystem 
becomes a single site with Hubbard repulsion $U$ at the end point of a semi-infinite 
lattice. This equivalent Anderson model is sketched in Fig.~\ref{FIGURE2}. 
To show this mapping, we define the fermion operators 
\begin{align}
a_{\mathrm e,x}^{\dagger}& = (c_{-x+1}^{\dagger} + c_x^{\dagger})/{\sqrt 2},\\
a_{\mathrm o,x}^{\dagger}& = (c_{-x+1}^{\dagger} - c_x^{\dagger})/{\sqrt 2}\,,
\end{align}
which create a spinless fermion in an even/odd ($\mathrm e/\mathrm o$) combination 
of the orbitals located at the sites $x$ and $-x+1$ of the original infinite lattice, 
(or a fermion with pseudo-spin $\sigma=\mathrm e/ \mathrm  o$ in the transformed 
semi-infinite lattice). $a_{\mathrm e/\mathrm o,x}$ are the corresponding annihilation 
operators. Expressing $H_{ns}$ in terms of these new operators, one gets 
\begin{equation}
H_{ns} = (V_G - t_d) n_{\mathrm e} + 
(V_G + t_d) n_{\mathrm o} + U n_{\mathrm e} n_{\mathrm o}\,, 
\label{Anderson}
\end{equation}
where $n_{\sigma}=a_{\sigma,1}^{\dagger} 
a_{\sigma,1}^{\vphantom{\dagger}}$ and where the pseudo-spin ``$\mathrm  e$'' 
(``$\mathrm o$'') is parallel (anti-parallel) to the ``Zeeman field'' $t_d$. 
In terms of the operators 
\begin{equation}
d^{\dagger}_{k,\sigma}=\sqrt{2/\pi} \sum_{x=2}^{\infty} 
\sin( k (x-1)) a^{\dagger}_{\sigma,x}
\end{equation}
creating a spinless fermion of 
pseudo-spin $\sigma$ and momentum $k$ in the transformed semi-infinite 1D-lead, 
the lead and the coupling Hamiltonians can be written as
\begin{equation}
H_{l} = \sum_{k,\sigma} \epsilon_k n_{k,\sigma} 
\end{equation}
and 
\begin{equation}
H_{c} = \sum_{k,\sigma} V (k) ( a_{\sigma,1}^\dagger 
d_{k,\sigma}^{\vphantom\dagger}+ H. c. )\,, 
\end{equation}
where the $k$-dependent hybridization 
\begin{equation}
V(k)=-t_c \sqrt{2/\pi} \sin k
\end{equation}
yields an impurity level width at $E_F$ which is given by
\begin{equation}
\Gamma=\frac{t_c^2}{t_h} \sin k_F\,,
\label{gamma}
\end{equation}
$n_{k,\sigma}= d_{k,\sigma}^\dagger d_{k,\sigma}^{\vphantom\dagger}$
and $\epsilon_k=-2t_h \cos k$.
 
One can see that ISIM is identical to an Anderson model with a local magnetic field 
$t_d$ which acts on the impurity only and gives rise to the Zeeman terms $\pm t_d$ in 
Eq.~\eqref{Anderson}. Therefore, in the limit $t_d \to 0$, ISIM must exhibit an orbital 
Kondo effect if the equivalent Anderson model can be reduced to a Kondo model. The 
fact that the impurity is not coupled to a 3D bath of conduction electrons, but only 
to a single semi-infinite 1D bath changes only the proportionality factor of the hybridization 
function. We underline that the dimensionality of the considered baths of conduction electrons 
does not play a significant role in Kondo physics, such that the results of this study should hold 
if one attaches 2D bars or 3D strips instead of 1D leads to the same nanosystem.

\section{Corresponding NRG Chain} 
\label{NRG Chain}

ISIM can be studied using the NRG procedure \cite{krishna-murthy2,bulla,hewson} developed 
by Wilson for the Anderson model after minor changes. First, we assume 
$V(k) \approx V(k_F=\pi/2)$ and, using standard NRG procedure, we divide the conduction 
band of the electron bath into logarithmic sub-bands characterized by an index $n$ and 
an energy width 
\begin{equation}
d_n=\Lambda^{-n}(1-\Lambda^{-1})\,.
\end{equation}
Throughout this paper, we use the discretization parameter $\Lambda = 2$. 
Within each sub-band, we introduce a complete set of orthonormal functions 
$\psi_{np} (\epsilon)$, and expand the lead operators in this basis. Dropping the 
terms with $p \neq 0$ and using a Gram-Schmidt procedure, the original 1D leads 
give rise to another semi-infinite chain with nearest-neighbor hopping
terms, each site $n$ representing now a conduction electron excitation at 
a length scale $\Lambda^{n/2} k_F^{-1}$ centered at the impurity. In this transformed 
1D model shown in Fig.~\ref{FIGURE3} and hereafter called the NRG chain, 
the impurity and the $N-1$ first sites form a finite chain of length $N$, which is 
described by the Hamiltonian $H_N$, the successive sites $n$ and $n+1$ being coupled 
by hopping terms $x_{n}$ which decay exponentially as $n \to \infty$ and are given by:
\begin{equation}
x_{n} =\Lambda^{-n/2} 
\frac{(1+\Lambda^{-1})(1-\Lambda^{-n-1})}
{2\sqrt{(1-\Lambda^{-2N-1})(1-\Lambda^{-2N-3})}}\,.
\label{hopping_NRG}
\end{equation}
The impurity is coupled to the first site of the NRG chain by an hopping term
\begin{equation}
t'_c = \frac{t_c}{(8\pi^3)^{1/4}}
\left(\log\left(\frac{\Lambda(\Lambda+1)}{\Lambda-1}\right)\right)^2\,.
\label{coupling_NRG}
\end{equation}
Since the length $N$ is related \cite{krishna-murthy2} to the temperature $T$ by 
the relation
\begin{equation}
k_BT \approx \frac{1+\Lambda^{-1}}{2}\Lambda^{-(N-1)/2}\,,
\end{equation}
$N$ can be interpreted as a logarithmic temperature scale ($N \propto -\log T$), the 
large values of $N$ corresponding to temperatures $T$ small compared to the bandwidth 
$t_h$.  

The NRG chain coupled to the impurity is iteratively diagonalized and rescaled, 
the spectrum being truncated to the $N_s$ first states at each iteration (We use $N_s=1024$ 
in this study). The behavior of ISIM as $T$ decreases can be obtained from the spectrum of 
$H_N$ as $N$ increases, the bandwidth of $H_N$ being suitably rescaled at each step.  A fixed 
point of the renormalization group (RG) flow corresponds to an interval of successive iterations $N$ of the same parity, 
where the rescaled many-body excitations $E_{I}(N)$ do not vary. The fixed point is therefore 
characterized inside this interval by two spectra, one characterizing the even values of $N$, 
the other the odd values. If it is a free fermion fixed point, 
$E_{I}=\sum_{\alpha} \epsilon_{\alpha}$, the $\epsilon_{\alpha}$ being effective one-body excitations, 
and the interacting system behaves as a non-interacting system with renormalized parameters 
$\widetilde{ t_d}$ and $\widetilde{t_c}$ near the fixed point.  Moreover, if one has free fermions 
when $T \to 0$, the conductance $g$ can be directly extracted from the NRG spectrum. 

\section{Restriction to the symmetric case} 
\label{symmetric case}
Using the NRG procedure, ISIM can be studied as a function of $T$ for 
arbitrary values of its 6 bare parameters $U$, $E_F$, $V_G$, $t_d$, 
$t_c$ and $t_h$. Hereafter, we take $E_F=0$ and $V_G=-U/2$. This choice 
makes ISIM invariant under particle-hole symmetry, with a uniform density 
$\langle n_x\rangle=1/2$. Very often, the infinite bandwidth 
limit ($t_h \to \infty$) is assumed in the theory of quantum impurities. 
This corresponds to magnetic alloys where the bandwidth of the conduction 
electrons is large compared to the other energy scales of the model. In 
this study, we take $t_h=1$, which defines the energy scale and allows us 
to consider also mesoscopic regimes where the scales $U$, $t_d$, $t_c$ or 
$V_G$ can exceed $t_h$. In that case, Eq.~\eqref{gamma} gives for the levels 
$\pm t_d$ of the isolated non-interacting nanosystem a width 
\begin{equation}
\Gamma=t_c^2 
\end{equation}
when the nanosystem is coupled to leads. Our motivation to restrict the study to 
the symmetric case is not justified by physical considerations, but mainly for the 
sake of simplicity, restricting the RG flow into a space of 3 effective parameters 
$( {\tilde U}, \widetilde{t_c}, \widetilde{t_d})$ only. Doing so, we proceed 
as Krishna-murthy, Wilkins and Wilson for the Anderson model, studying first the 
symmetric case \cite{krishna-murthy2} before considering later the asymmetric case 
\cite{krishna-murthy3} and its characteristic valence-fluctuation regime.
\begin{figure}
\centerline{
\includegraphics[width=0.8\columnwidth]{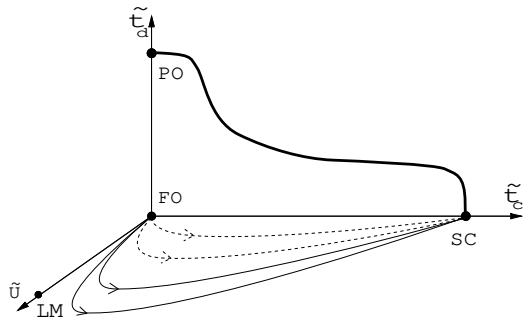}
}
\caption
{Line of free fermion fixed points [thick solid line in the plane 
$(\widetilde{t_d},\widetilde{t_c})$], characterizing ISIM 
when $T \to 0$ as $t_d$ increases from $t_d=0$ (SC fixed point) towards 
$t_d \to \infty$ (PO fixed point). The FO, LM and SC free fermion fixed 
points and the RG trajectories \cite{hewson} followed by ISIM as $T$ decreases 
for $t_d=0$ are indicated in the plane $(\widetilde{U},\widetilde{t_c})$, 
for $\pi \Gamma > U$ (dashed) and $\pi \Gamma <U $ (solid).
\label{FIGURE4} 
}  
\end{figure}
\begin{figure}
\includegraphics[width=\columnwidth]{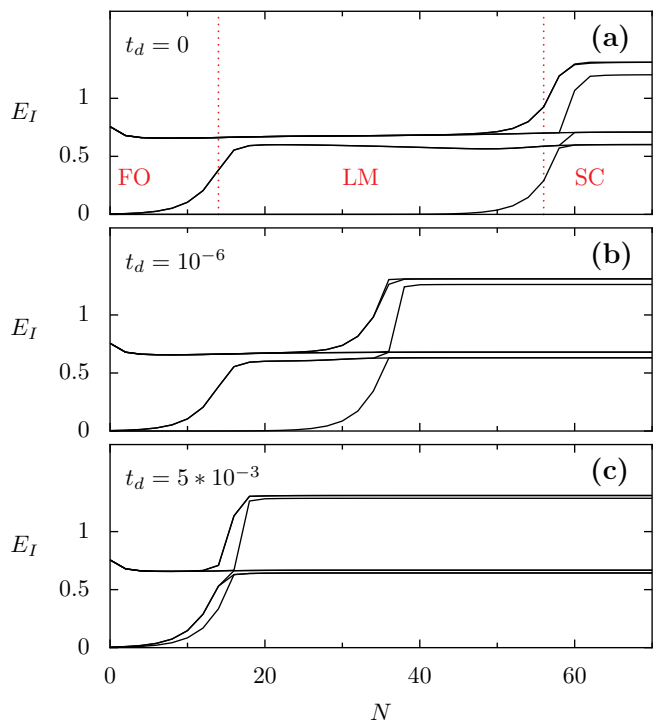}
\caption{(Color online) Many body excitations $E_{I}$ as a function of $N$ (even values) 
for  $U=0.005$ and $t_c=0.01$. For $t_d=0$ (Fig.~\ref{FIGURE5}a), one can see 
the 3 successive plateaus (FO, LM and SC fixed points) of the Anderson model.
As $t_d$ increases (Fig.~\ref{FIGURE5}b and Fig.~\ref{FIGURE5}c), the LM 
plateau shrinks and disappears when $t_d \gg t_d^*=t_c\sqrt{U}$. 
\label{FIGURE5}
}
\end{figure}

\section{Role of the temperature \boldmath{$T$}}
\label{Three fixed points}

When $t_d=0$, ISIM is an Anderson model which has the RG flow 
sketched in Fig.~\ref{FIGURE4} for the particle-hole symmetric case. 
At low  values of $N$ (high values of $T$), ISIM is located in 
the vicinity of the unstable free-orbital (FO) fixed point. As 
$N$ increases ($T$ decreases), ISIM flows towards the stable 
strong-coupling (SC) fixed point. 

If the interaction is weak ($U < \pi \Gamma$), its effects can be 
described by perturbation theory, the flow goes 
directly from the FO fixed point towards the SC fixed point, and 
there appears no orbital Kondo effect for $t_d \to 0$. 

If the interaction is large ($U > \pi \Gamma$), the flow can visit an  
intermediate unstable fixed point---the local-moment (LM) fixed 
point---before reaching the SC fixed point. In that case, ISIM is identical to a 
Kondo model characterized by a temperature $T_K$ and by universal 
functions of the ratio $T/T_K$. For $t_d=0$, ISIM is on the FO fixed 
point when $t_c \sqrt{U} < T$, exhibits a local moment when 
$T_K< T  < t_c\sqrt{U}$  and reaches the SC fixed point when $T < T_K$. 
While the Hartree-Fock theory qualitatively describes \cite{anderson} the 
local moment at high temperatures $T_K < T < t_c\sqrt{U}$, it breaks down 
at low temperatures ($T < T_K$), where the effect of the interaction 
becomes non-perturbative and gives an orbital Kondo effect.

In Fig.~\ref{FIGURE5}, the first many-body excitations $E_{I}$ of ISIM 
are given for increasing even values of $N$ for $t_d=0$. Since $U > \pi t_c^2$, 
one gets 3 plateaus corresponding to the 3 expected fixed points. Inside the 
plateaus, the spectra are free fermion spectra which are described in 
Ref.~\onlinecite{krishna-murthy2}. However, between the plateaus, there are 
no free fermion spectra and $E_{I}\neq \sum_{\alpha} \epsilon_{\alpha}$. 
As $t_d$ increases (Fig.~\ref{FIGURE5}), the LM plateau decreases and 
vanishes when $t_d$ reaches a value $\approx t_c \sqrt{U}$. 

\begin{figure}
\centerline{
\includegraphics[width=0.8\columnwidth]{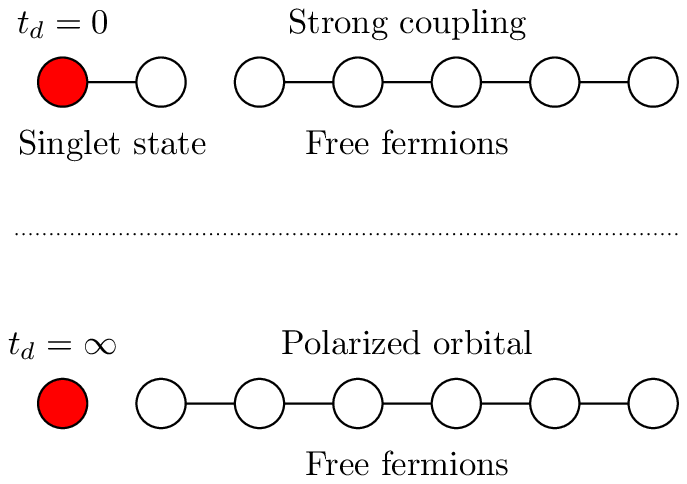}
}
\caption
{(Color online) NRG chain in the SC limit (upper figure) and in the PO limit (lower figure). 
When $t_d \to 0$, the impurity (red dot) and the first site of the NRG chain 
form a system in its singlet ground state decoupled from the other sites which 
carry free fermion excitations. When $t_d \to \infty$, the even (odd) orbital 
of the impurity  is occupied (empty) and the other sites carry free fermion excitations. 
Therefore, there is a permutation of the parity of the length of the free part as $t_d$ 
increases. When the excitations of the free part are independent of this parity, $t_d=\tau$ 
and $g=1$.
\label{FIGURE6} 
}  
\end{figure}
\begin{figure}
\includegraphics[width=\columnwidth]{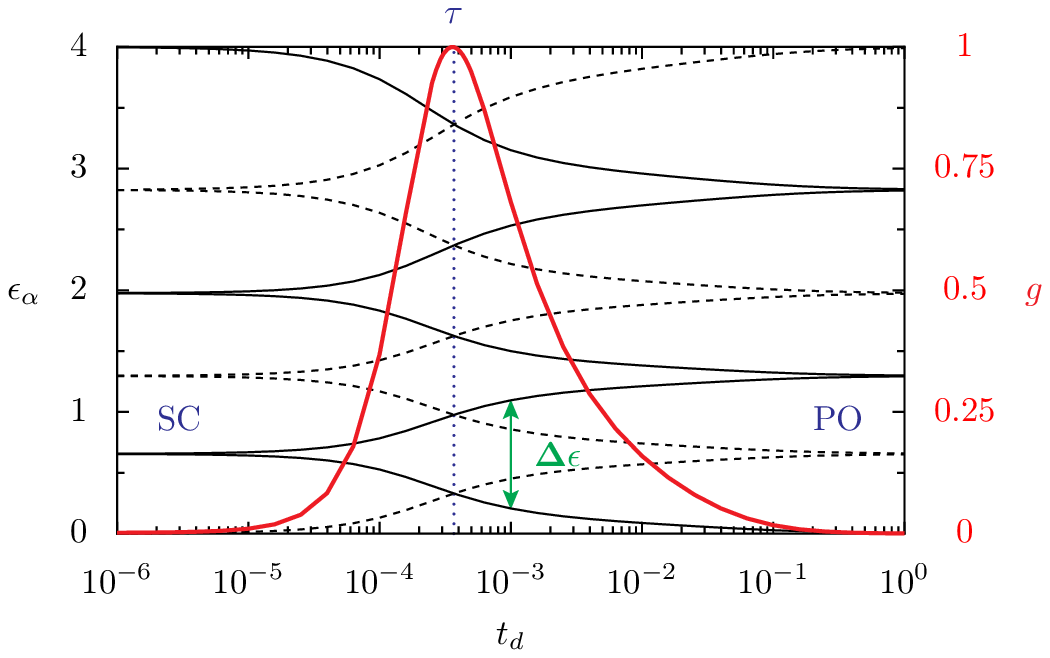}
\caption{(Color online) One body excitations $\epsilon_{\alpha} (t_d)$ [extracted 
from the $E_{I}(N \to \infty, t_d)$] for $U=0.1$ and $t_c=0.1$ (left scale). 
The solid (dashed) lines correspond to NRG chains of even (odd) length $N$. 
Conductance $g(t_d)$ extracted from $\Delta \epsilon(t_d)$ using 
Eq.~\eqref{transmission} (thick red curve, right scale). 
For $t_d=\tau$, the $\epsilon_{\alpha}$ are independent of the parity of $N$ and $g=1$.   
\label{FIGURE7}
}
\end{figure}

\section{Role of the internal hopping at \boldmath{$T=0$}} 
\label{free fermion fixed points}

Let us study how the many-body levels $E_I$ given by the NRG algorithm 
depend on the internal hopping term $t_d$ in the limit where $N \to \infty$, 
i.e.\ in the limit where the temperature $T \to 0$. 

When $t_d=0$, one has the SC limit \cite{krishna-murthy2} of the Anderson model 
where the impurity is strongly coupled to its first neighbor in the NRG chain 
(the conduction-electron state at the impurity site). The impurity and this site 
form a system which can be reduced to its ground state (a singlet), the $N-2$ other 
sites carrying free fermion excitations $\epsilon_{\alpha}$ which are independent 
of that interacting system. This SC limit of the Anderson model without field is 
sketched in Fig.~\ref{FIGURE6} (upper part). In the Kondo model, the site directly 
coupled to the impurity describes the cloud of conduction electrons which fully 
screens the magnetic moment of the impurity.  

When $t_d \to \infty$, the impurity occupation numbers $\langle n_{\mathrm e}\rangle \to 1$ 
and $\langle n_{\mathrm o}\rangle \to 0$, and the $N-1$ other sites of the NRG chain are 
independent of the impurity. This limit is sketched in Fig.~\ref{FIGURE6} (lower part).
We call this limit ``Polarized  Orbital (PO),'' since it coincides with the 
FO fixed point of the Anderson model, except that the spin of the free orbital is not free, 
but fully polarized in our case. 

The $E_{I}(t_d)$ correspond to many-body excitations of effective non-interacting 
spectra when $t_d\to 0$ and $t_d \to \infty$. When $t_d$ varies between those 2 limits, 
the NRG algorithm continues to give many-body excitations $E_{I}(t_d)$ compatible with 
the free fermion rule $E_{I}(t_d)=\sum_{\alpha} \epsilon_{\alpha} (t_d)$, allowing us to 
extract one-body excitations $\epsilon_{\alpha}(t_d)$ for intermediate values of $t_d$.  
We conclude that there is a continuum of effective non-interacting spectra which describe 
the $E_{I}$ as $t_d$ varies in the limit $N \to \infty$, i.e.\ the $T \to 0$ limit of ISIM 
is given by a continuum line of free fermion fixed points. This line is sketched in 
Fig.~\ref{FIGURE4}. Having always free fermions as $t_d$ varies means that the $T=0$ 
scattering properties of an interacting region embedded inside an infinite non-interacting 
lattice are those of an effective non-interacting system with renormalized parameters, in 
agreement with the DMRG study of the persistent current given in Ref.~\onlinecite{molina1}. 
We underline that those effective non-interacting spectra describe the $T=0$ limit, while a 
description of the low-temperature dependence of the conductance requires effective Hamiltonians 
of Landau quasiparticles with residual quasiparticle interactions. Such a Fermi 
liquid theory has been proposed by Nozi\`eres. In the case of the Anderson model, it has been 
developed in Ref.~\onlinecite{hewson1} without magnetic field ($t_d=0$) and in Ref.~\onlinecite{hewson2} 
with magnetic field ($t_d \neq 0$). However, the quasiparticle interaction comes into 
play at finite temperatures only and a residual interaction is not necessary for describing the 
${E_{I}}$ as a function of $t_d$ in the limit $N \to \infty$. 

Fig.~\ref{FIGURE7} shows these first one-body excitations $\epsilon_{\alpha}$ as a function of $t_d$ 
extracted from the $E_{I}(t_d)$, calculated with $U=0.1$ and $t_c=0.1$. The pseudo-spin degeneracy 
being broken by the ``magnetic field'' $t_d \neq 0$, the first (second) one-body excitation 
$\epsilon_1$ ($\epsilon_2$) carries respectively an even (odd) pseudo-spin if $N$ is even. This is 
the inverse if $N$ is odd, $\epsilon_1$ ($\epsilon_2$) carrying respectively an odd (even) pseudo-spin. 

\section{Perfect Transmission and Characteristic Energy Scale}
\label{Characteristic Energy Scale}

Since the free part of the NRG chain has $N-2$ sites for $N \to \infty$ and $t_d \to 0$ 
(SC fixed point), while it has $N-1$ sites for $t_d \to \infty$ (PO fixed point), there is 
a permutation of the $\epsilon_{\alpha}(t_d)$ as $t_d$ increases: the $\epsilon_{\alpha}
(t_d\to 0)$ for $N$ even become the $\epsilon_{\alpha}(t_d \to \infty)$ for $N$ odd and 
vice-versa. This permutation is shown in Fig.~\ref{FIGURE7}.
Since for $N \to \infty$ there is a permutation between the even and odd spectra 
as $t_d$ increases, there is a value of $t_d$ for which the 
$\epsilon_{\alpha} (t_d)$ are independent of the parity of $N$. This 
value defines very precisely the characteristic energy scale $\tau (U,t_c)$ of ISIM. 
Because of particle-hole symmetry, the nanosystem (the impurity of the 
NRG chain) is always occupied by one electron. Binding one electron of 
the leads with this electron reduces the energy when $t_d<\tau$, while 
it increases the energy when $t_d > \tau$. For $t_d=\tau$, it is indifferent 
to bind or not an electron of the lead with the one of the nanosystem, making 
ISIM perfectly transparent. This gives the proof that, for all values 
of $U$ and $t_c$, there is always a value $\tau$ of $t_d$ for which the 
interacting region becomes perfectly transmitting and 
\begin{equation}
g(t_d=\tau(U,t_c))=1\,. 
\end{equation}
The argument is reminiscent to that giving the condition for having a perfectly 
transparent quantum dot in the Coulomb blockade regime: $t_d$ in our case, the gate 
voltage in the other case, have to be adjusted to values for which it costs 
the same energy to put an extra electron outside or inside the dot. 
The quantum conductance $g$ can be extracted from the NRG spectra. Using a method 
explained in the following section, we have calculated $g(t_d)$. The result shown 
in Fig.~\ref{FIGURE7} confirms that $g=1$ precisely for the value $\tau$ of $t_d$ 
for which the $N\to \infty$ low energy excitations are independent of the parity of 
$N$.

\section{Free-fermion spectra and quantum conductance}
\label{quantum conductance}
\begin{figure}
\centerline{
\includegraphics[width=\columnwidth]{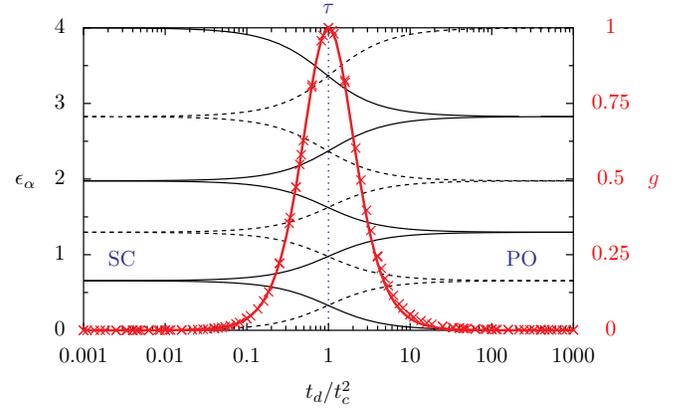}
}
\caption{(Color online) $\epsilon_{\alpha}$ and $g$ as a function of $t_d/t_c^2$ 
without interaction ($U=0$). The values of $g$ extracted from the NRG spectra (red cross) 
coincide with the exact values (Eq.~\eqref{conductance U=0} - red line). 
\label{FIGURE8}
}
\end{figure}

As pointed out in previous works \cite{borda,hofstetter,oguri}, the quantum conductance 
$g$ can be directly extracted from the NRG spectra. Let us consider an NRG chain of even 
length $N \to \infty$. When $t_d=0$, the one body spectrum is identical for the 
two pseudo-spins. A hopping $t_d \neq 0$ breaks the pseudo-spin degeneracy and 
opens a gap $\Delta \epsilon(t_d)$ (indicated in Fig.~\ref{FIGURE7}) between excitations 
of opposite pseudo-spins. For free fermions in the limit $T \to 0$, the asymptotic 
electron states of ISIM are stationary waves with even (odd) phase shifts $\delta_{\mathrm e,\mathrm o}(k)$ 
induced by the scattering region  
\begin{align}
\Psi_{\mathrm e}(k,p) &\propto \cos(k(p-1/2)-\delta_{\mathrm e}(k)), \\
\Psi_{\mathrm o}(k,p) &\propto \sin(k(p-1/2)-\delta_{\mathrm o}(k))\,.
\end{align}
The shifts $\delta_{\mathrm e,\mathrm o}(k)$ of the scattering phases and $\delta \epsilon_{\mathrm e,\mathrm o}(k)$ of 
the energy levels  are proportional for each pseudo-spin. This can be shown by taking 
a finite size $L$ for ISIM, quantizing the momenta $k_{\mathrm e,\mathrm o}(n)=\pi n/L +\delta_{\mathrm e,\mathrm o}(k)/L$ 
and using the dispersion relation $\epsilon(k)=2\cos(k)$. This yields   
\begin{equation}
\Delta \epsilon= (\delta \epsilon_{\mathrm e}- \delta \epsilon_{\mathrm o}) \propto  (\delta k_{\mathrm e}- \delta k_{\mathrm o}) 
\propto (\delta_{\mathrm e} - \delta_{\mathrm o})\,.
\label{exact_relation}
\end{equation}
In the limit $L \to \infty$, the quantum conductance  $g(t_d)$ of ISIM 
can be expressed as a function of the scattering phase shifts:  
\begin{equation}
g(t_d)=\sin^2(\delta_{\mathrm e}(t_d) - \delta_{\mathrm o}(t_d))\,.
\label{phaseshift}
\end{equation}
From Eq.~\eqref{exact_relation} and Eq.~\eqref{phaseshift}, one eventually obtains 
the relation which allows us to extract $g$ from the NRG spectra: 
\begin{equation}
g(t_d)=\sin^2 \left(\pi \frac{\Delta \epsilon(t_d)}
{\Delta\epsilon(t_d \to \infty)}\right)\,. 
\label{transmission} 
\end{equation}
The proportionality factor between $\Delta \epsilon(t_d)$ and $\delta_{\mathrm e}(t_d) - 
\delta_{\mathrm o}(t_d)$ has been determined from the condition that $g \to 0$ (for $\delta_{\mathrm e} - \delta_{\mathrm o} 
\to \pi$) when $t_d \to \infty$. Eq.~\eqref{transmission} describes a quantum 
conductance which vanishes when $t_d \to 0$ and when $t_d \to \infty$, and reaches 
the unitary limit for an intermediate value $\tau(U,t_c)$ of $t_d$. 

 Let us check that the NRG algorithm and Eq.~\eqref{transmission} give us the correct 
behavior for the conductance $g$ in the non-interacting limit where it is straightforward 
to solve the scattering problem. One obtains  
\begin{equation}
g(U=0)= 4 \left(\frac {t_d}{t_c^2}+\frac {t_c^2}{t_d} \right)^{-2}
\label{conductance U=0}\,.
\end{equation}
The $\epsilon_{\alpha}(t_d)$ given by the NRG algorithm for $U=0$ are shown in 
Fig.~\ref{FIGURE8} with the corresponding values of $g$ obtained from  
Eq.~\eqref{transmission}. One can see that the behavior of $g$ given 
by the NRG algorithm and Eq.~\eqref{transmission} reproduces the correct 
behavior given by Eq.~\eqref{conductance U=0} in the non-interacting limit. 

\section{Universal properties of ISIM} 
\label{Universality}
\begin{figure}
\includegraphics[width=\columnwidth]{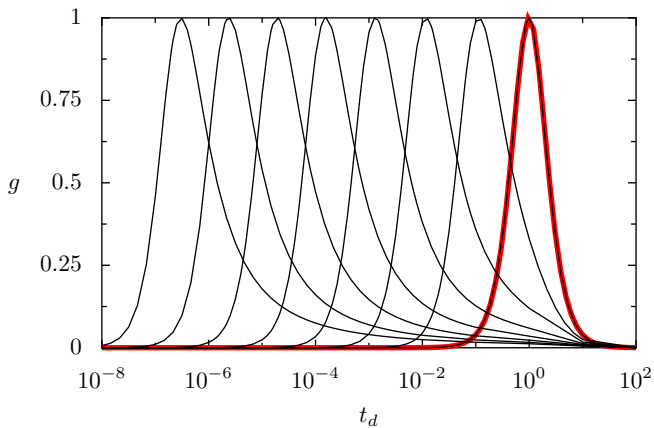}
\caption{(Color online) 
$g(t_d)$ extracted from the NRG spectra for $t_c=1$ and many values of $U$. 
Increasing $U$ shifts the transmission peaks to smaller values of $t_d$. The 
curves correspond respectively to $U=0$ (red curve), $5, 10, 15, 20, 25, 30, 
35$. 
\label{FIGURE9}
}
\end{figure}
\begin{figure*}
\includegraphics[width=2\columnwidth]{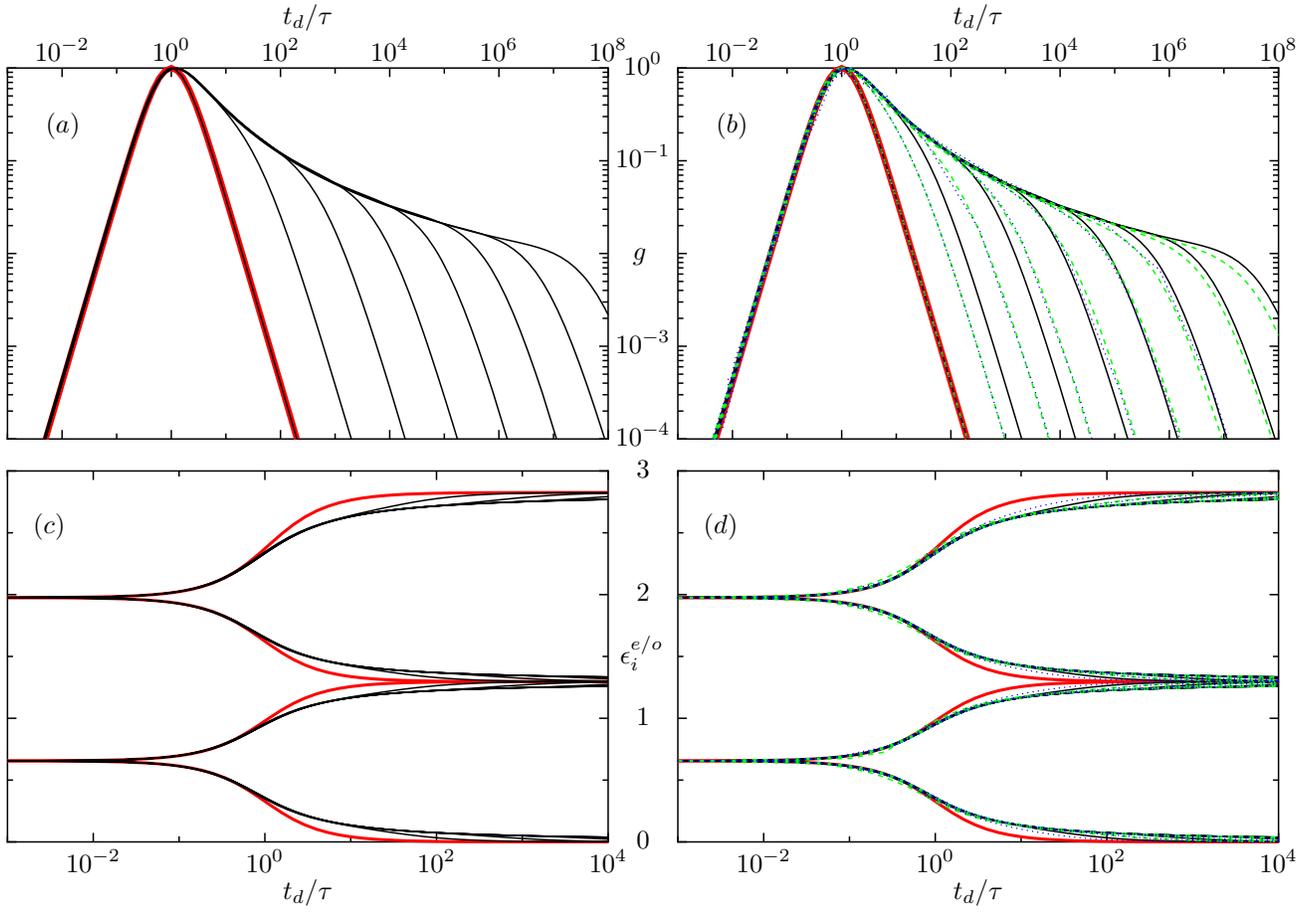}
\caption{(Color online) 
Figure~\ref{FIGURE10}(a): $g$ as a function of $t_d/\tau$ for $t_c=1$ and 
many values of $U$. $\tau$ has been determined from the criterion 
$g(t_d=\tau)=1$. The larger is $U$, the larger are the values of 
$t_d/\tau$ where $g$ decays. The curves correspond respectively to 
$U=0$ (red curves) and $U=5, 10, 15, 20, 25, 30, 35$ (black curves). 
Figure~\ref{FIGURE10}(b): $g$ as a function of $t_d/\tau$. To the data 
(black curves) calculated taking $t_c=1$ and shown in Fig.~\ref{FIGURE10}(a), 
we have added the data calculated taking $t_c=0.1$ and $U=0$ (red curve), 
$0.05, 0.1, 0.15, 0.2, 0.25, 0.3, 0.35, 0.4$ (green dashed curves) 
and taking $t_c=0.01$ and $U=0$ (red curve), $0.0005, 0.001, 0.0015, 0.002, 
0.0025, 0.003, 0.0035$ (blue dotted curves). Figure~\ref{FIGURE10}(c): 
First one body excitations $\epsilon_{\alpha}$ for $N$ even as a function of 
$t_d/\tau$. Same values of $t_c$ and $U$ as in Fig.~\ref{FIGURE10}(a).
Figure~\ref{FIGURE10}(d): $\epsilon_{\alpha}$ for $N$ even as a function of 
$t_d/\tau$. Same values of $t_c$ and $U$ as in Fig.~\ref{FIGURE10}(b).
\label{FIGURE10}
}
\end{figure*} 
\begin{figure}
\centerline{
\includegraphics[width=\columnwidth]{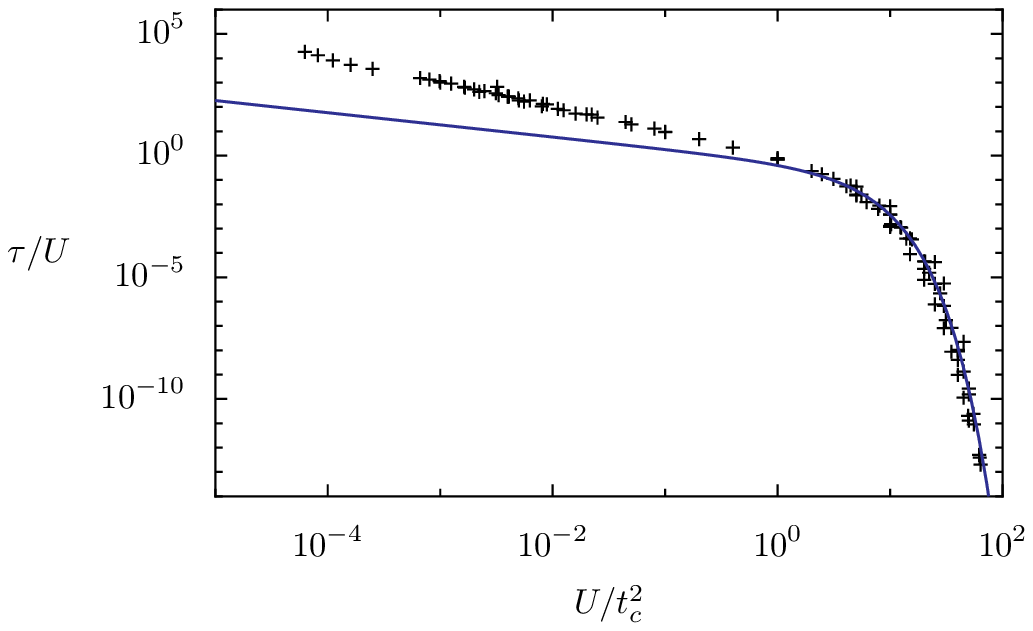}
}
\caption{(Color online) Characteristic scale $\tau(U,t_c)$ as a function of $U$ 
and $t_c$. The values of $\tau(U,t_c)/U$ obtained from the condition $g(t_d/\tau)=1$ 
and used in Fig.~\ref{FIGURE10} are plotted as a function of $x=U/t_c^2$ (+). 
The solid blue line $y(x)= 0.728 \sqrt{2/(\pi x)} \exp-(\pi x/8)$ fits the 
data in the non-perturbative regime and corresponds to the relation $\tau=2 T_K$ 
with $T_k=0.364 \sqrt{2t_c^2U/\pi} \exp -(\pi U/ 8 t_c^2)$.
}  
\label{FIGURE11} 
\end{figure}
 In Fig.~\ref{FIGURE9}, the conductance $g$ extracted from the NRG 
spectra using Eq.~\eqref{transmission} is given as a function of $t_d$ 
for a coupling term $t_c=1$ and many values of $U$. The larger is $U$, 
the smaller is the characteristic scale $\tau(U)$. Fig.~\ref{FIGURE9} 
seems to indicate that the left sides of the transmission peaks are simply 
translated to lower values of $t_d$ as $U$ increases. This is confirmed in 
Fig.~\ref{FIGURE10}a) where $g$ is given as a function of the dimensionless 
scale $t_d/\tau$, $\tau$ being obtained from the criterion $g(t_d=\tau)=1$. 
The curves $g(t_d/\tau)$ obtained for $t_c=1$ are shown in Fig.~\ref{FIGURE10}a), 
while the curves obtained for $t_c=0.1$ and $t_c=0.01$ have been added in 
Fig.~\ref{FIGURE10}(b). These figures show the main result of this study, i.e., 
When $t_d$ is not too large, $g$ is given by a universal function of $t_d/\tau(U,t_c)$. 
This function is independent of the values taken for $U$ or $t_c$. Since $g(t_d)$ has been 
directly extracted from the free fermion NRG spectra, the NRG spectra must be 
also given by universal functions of $t_d/\tau(U,t_c)$, independent of the 
values of $U$ and $t_c$. This is shown in Fig.~\ref{FIGURE10}(c) ($t_c=1$) 
and in Fig.~\ref{FIGURE10}(d) ($t_c=1,0.1,0.01$), where the first excitations 
$\epsilon_{\alpha}$ obtained for NRG chains of large even length $N$ are plotted 
as a function of $t_d/\tau$. 

The values of $\tau(U,t_c)$ used in Fig.~\ref{FIGURE10} are given in Fig.~\ref{FIGURE11}. 
If one decreases the temperature $T$ in the Anderson model without magnetic field, it has been 
shown in Ref.~\onlinecite{krishna-murthy2} that the interaction effects remain perturbative when 
$U < \pi \Gamma$ while they become non-perturbative when $U > \pi \Gamma$. Since the level width 
$\Gamma=t_c^2$ for the nanosystem used in ISIM at half-filling, we give the dimensionless scale 
$\tau/U$ as a function of $U/t_c^2$ in Fig.~\ref{FIGURE11}. One can see that $\tau/U$ has a slow 
decay followed by a faster decay, with a crossover around an interaction threshold consistent 
with the interaction  threshold $\pi \Gamma$ characterizing the perturbative-non-perturbative 
crossover in the Anderson model. For ISIM, this suggests that the interaction effects upon 
$g$ are perturbative when $U < \pi \Gamma$, and non-perturbative when $U > \pi \Gamma$. However, 
the universal behavior of $g(t_d/\tau)$ is not restricted to the non-perturbative ``Kondo'' regime, 
but characterizes also the perturbative regime.

\subsection{Universality near the SC limit (\boldmath{$t_d \le \tau$})}
\label{SC limit}
The universal regime can be divided into two parts. The first one begins at 
$t_d=0$ in the vicinity of the SC fixed point and ends at $t_d = \tau$. 
In this SC regime, $g$ behaves as without interaction, but with a renormalized 
level width ($t_c^2 \to \tau(U,t_c)$). As shown in Fig.~\ref{FIGURE10}, all the 
curves $g(t_d/\tau)$ are on a single universal curve when $t_d \le \tau$, 
independent of the values of $U$ and $t_c$. Since one of those curves 
(the red one) corresponds to the non-interacting limit $U=0$, the universal 
curve for $t_d \leq \tau$ is given by Eq.~\eqref{conductance U=0} where 
$t_d$ is measured in units of $\tau$, instead of $t_c^2$:
\begin{equation}
g(t_d,U,t_c)= 4 \left(\frac {t_d}{\tau(U,t_c)}+\frac {\tau(U,t_c)}{t_d} 
\right)^{-2}
\label{conductance Univ}\,.
\end{equation}

\subsection{Universality around a LM limit (\boldmath{$\tau < t_d < t_c \sqrt{U}$})}
\label{LM limit}
The universal regime persists when $t_d$ exceeds $\tau$. While $g(t_d/\tau)$ decays 
immediately after the transmission peak if $U=0$, $g(t_d/\tau)$ begins to follow 
a new part of the universal curve if $U \neq 0$. This new part is not given by 
Eq.~\eqref{conductance Univ}, and ceases when a faster decay occurs. The larger is $U$, 
the larger is the interval of values of $t_d/\tau$ where $g(t_d/\tau)$ follows the 
slow decay of the universal curve (see Fig.~\ref{FIGURE10}). To describe this slow decay, 
one can use exact results which give the magnetization $m(h)$ of the Anderson model at $T=0$ 
as a function of the magnetic field $h$. This will be done after a study of the relation 
between $g(t_d)$ and $m(h)$. 

Let us just note now that the singlet state of the SC limit could be broken either if the temperature 
$T$ or the ``Zeeman energy'' $t_d$ exceeds the Kondo temperature $T_K$. This makes likely that the 
effects of $T$ and $t_d$ would be somewhat similar. If this is the case, the intermediate values of 
$t_d$ would be related to the formation of a local moment in the equivalent Anderson model of ISIM, 
and an intermediate LM regime would take place between the SC regime for low values of $t_d$ and 
the PO regime for large values of $t_d$. This classification is used in Ref.~\onlinecite{tsvelick-wiegmann1} 
for describing the effect of a magnetic field in the zero-temperature limit of the Anderson model. However, 
we can only refer to the SC, LM and PO regimes, and not to the SC, LM and PO fixed points. The RG flows of 
ISIM yielded by increasing $t_d$ at $T=0$ and by increasing $T$ at $t_d=0$ are very different. Increasing 
$T$ in ISIM at $t_d=0$ yields the 3 plateaus shown in Fig.~\ref{FIGURE5}, characteristic of 3 well-defined 
fixed points. There are no free fermions between the plateaus, since there are no $\epsilon_{\alpha} (T)$ 
such that $E_{I}(T)= \sum_{\alpha} \epsilon_{\alpha} (T)$ outside the plateaus. In contrast, Fig.~\ref{FIGURE10} 
does not exhibit plateaus and the $E_{I}(t_d)$ can be described by a continuum of effective non-interacting 
spectra as  $t_d$ varies at $T=0$, and not only by the three spectra of the SC, LM and FO fixed points. 

\subsection{Interaction-independent conductance in the PO limit (\boldmath{$ t_d > t_c \sqrt{U}$})}
\label{PO limit}

\begin{figure}
\centerline{
\includegraphics[width=\columnwidth]{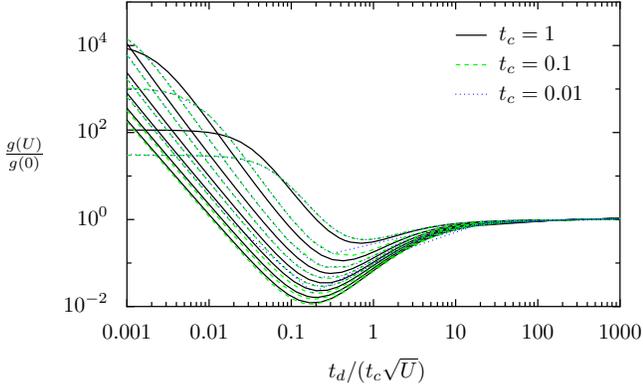}
}
\caption{(Color online)  Ratio $g(U,t_c,t_d)/g(U=0,t_c,t_d)$ as a function of $t_d/(t_c \sqrt{U})$. 
$g(U)$ is extracted from the NRG spectrum and $g(U=0)$ is given by Eq.~\eqref{conductance U=0}. One can see 
that $g$ behaves as without interaction when $t_d > t_d^* \approx 10 t_c \sqrt{U}$. Same values of 
$t_c$ and $U$ as in Fig.~\ref{FIGURE10}.
}  
\label{FIGURE12} 
\end{figure}

As one increases $t_d/\tau$, $g(t_d/\tau)$ eventually exhibits a fast decay which is not given by 
a universal function of $t_d/\tau$. As can be seen in Fig.~\ref{FIGURE10}, this fast decay corresponds 
to the decay of $g(t_d/t_c^2)$ obtained without interaction, but shifted to values of $t_d/\tau$ which 
increase when $U$ increases. According to Ref.~\onlinecite{tsvelick-wiegmann1}, the Anderson model at $T=0$ 
is in a LM regime if the magnetic field $h$ lies in the interval $T_K < h <h^* \propto \sqrt{U\Gamma}$. 
The upper threshold $\sqrt{U\Gamma}$ appears in the Hartree-Fock study made by Anderson~\cite{anderson} of the 
transition from the non-magnetic to the LM regime of the Anderson model, as one decreases $T$ without magnetic 
field. For ISIM, this suggests that the slow universal decay of $g(t_d/\tau)$ corresponding to the LM regime 
persists as far as $t_d < t_d^* \propto t_c \sqrt{U}$. Above $t_d^*$, ISIM should enter in the PO regime. 
This is confirmed in Fig.~\ref{FIGURE12}, where one can see that $g$ becomes independent of $U$ and behaves 
as without interaction when $t_d$ exceeds a threshold value $t_d^* \approx 10 \sqrt{U\Gamma}$.   

\section{Characteristic Energy Scale in the perturbative regime}
\label{Perturbative regime}

Without interaction, Eq.~\eqref{conductance U=0} implies that $g=1$ if $t_d=t_c^2$. 
This yields for the characteristic energy scale $\tau$ of ISIM a non-interacting value 
$t_c^2$. For weak values of $U$, there is a perturbative regime where $g$ 
and $\tau$ can be obtained using self-consistent Hartree-Fock theory. In the 
symmetric case, the Hartree corrections and the site potentials $V_G$ cancel each others 
since $V_G=-U/2$. The value of the inter-site hopping $t_d$ is modified 
because of exchange and takes \cite{asada} a value given by the self-consistent solution 
of the HF equation
\begin{equation}
v=t_d+U \langle c_0^\dagger c_1^{\vphantom{\dagger}} (v,t_c) \rangle\,.
\label{hopping-HF} 
\end{equation}
Using Eq.~\eqref{conductance U=0} with $v$ instead of $t_d$ gives the HF value  
of $g$.

 For the Anderson model, it is well known that HF theory fails to describe 
the Kondo regime. This Kondo regime occurs when the interaction exceeds a 
threshold value $\pi \Gamma$, either for low temperatures $T<T_K$ without 
magnetic field $h$, or for weak fields $h<T_K$ at $T=0$. In the frame of the 
HF approximation, a magnetic moment should be formed, while it actually 
vanishes because of strong correlations between the conduction electrons 
and the impurity spin (Kondo effect). Therefore, one does not expect that 
HF theory should be valid in the orbital Kondo regime of ISIM for large 
values of $U$ ($U > \pi \Gamma$) and small values of $t_d$ if $T=0$.  

\begin{figure}
\centerline{
\includegraphics[width=\columnwidth]{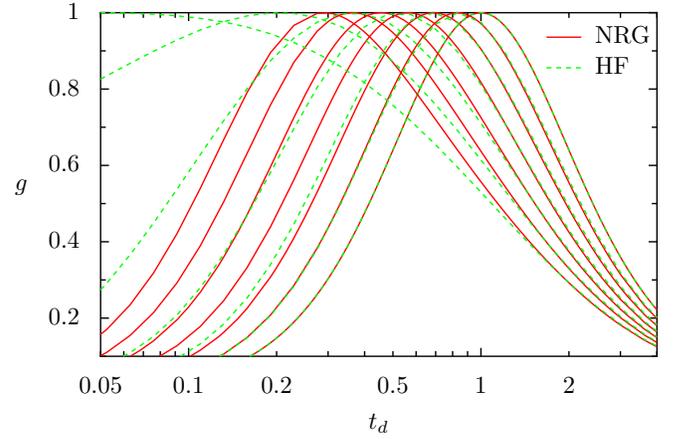}
}
\caption{(Color online) Conductance $g$ as a function of $t_d$ for $tc=1$ and 
$U=0,0.5,1,1.5,2,2.5,3$. $g$ extracted from the NRG spectra (solid red curves) 
coincides with the HF estimates (dashed green curves) when $U < t_c^2$ or 
for $t_d \gg t_c \sqrt{U}$ if $U > t_c^2$. 
}  
\label{FIGURE13} 
\end{figure}

\begin{figure}
\centerline{
\includegraphics[width=\columnwidth]{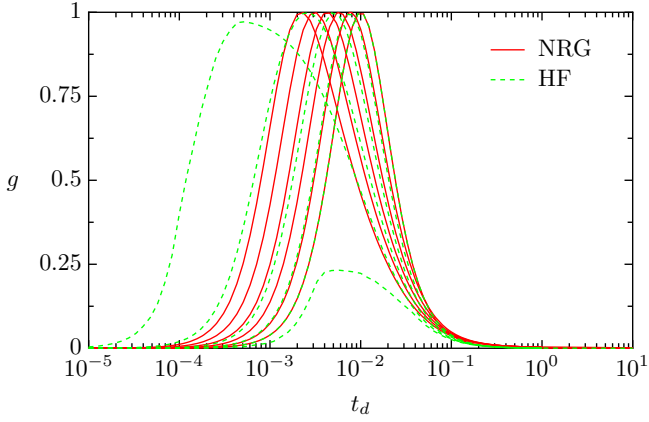}
}
\caption{(Color online) Conductance $g$ as a function of $t_d$ for $tc=0.1$ and 
$U=0, 0.01, 0.02, 0.03, 0.04$ and $U=0.05$. The NRG results (solid red curves) and
the HF results (dashed green curves) coincide for $U < t_c^2$. For $U=0.05$, 
the HF curve gives only a small peak where $g \approx 0.25$, and not $g=1$. 
} 
\label{FIGURE14} 
\end{figure}

For a large coupling  $t_c=1$, the results shown in Fig.~\ref{FIGURE13} confirm 
this prediction: HF theory gives the correct value of $g$ for all values of $t_d$ 
as far as $U$ remains smaller than $\pi t_c^2$. For $U > \pi t_c^2$, the HF curves 
and the NRG curves coincide only when $t_d$ is large ($t_d > \tau$), but 
become very different at the left side of the transmission peak. For a small 
coupling $t_c=0.1$, one can see in Fig.~\ref{FIGURE14} a more dramatic breakdown of 
HF theory which fails to give the peak of perfect transmission (see the curves with 
$U=0.04$ and more notably $U=0.05$ of Fig.~\ref{FIGURE14}).

\begin{figure}
\centerline{
\includegraphics[width=\columnwidth]{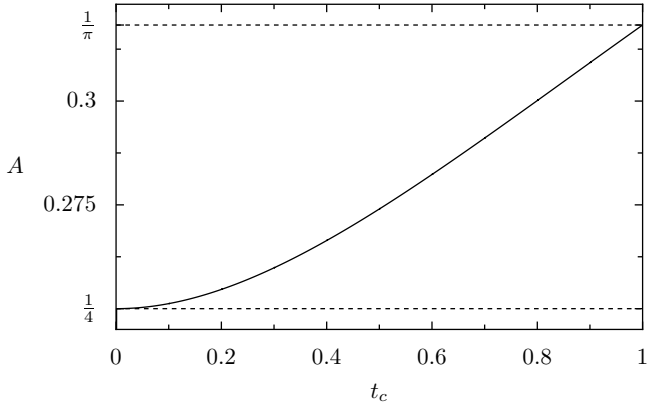}
}
\caption{Function $A(t_c)$ defined in Eq.~\eqref{FunctionA} as a 
function of $t_c$.
}  
\label{FIGURE15} 
\end{figure}

 The origin of this failure can be simply explained. In the perturbative regime 
where $g$ is given by Eq.~\eqref{conductance U=0} with $v$ instead of $t_d$, 
$g=1$ if $v=t_c^2$. This yields for the scale $\tau$ a HF value 
\begin{equation}
\tau_{HF}=t_c^2 - A(t_c) U,
\label{scale-HF} 
\end{equation}
where the function
\begin{equation}
A(t_c)=\langle c_0^\dagger c_1^{\vphantom{\dagger}} (v=t_c^2,t_c) \rangle\,,
\label{FunctionA} 
\end{equation}
which is shown in Fig.~\ref{FIGURE15}, depends weakly on $t_c$ ($A=1/\pi$ if $t_c=1$ 
while $A \to 1/4$ if $t_c \to 0$). When $U$ reaches a value $t_c^2/A$, 
HF theory predicts that the interaction should renormalize $t_d$ to a value $\tau_{HF}=0$ 
for having $g=1$! This is absurd and confirms that HF theory breaks down for ISIM above 
an interaction threshold which is essentially the same as for the Anderson model ($U > 
\pi \Gamma$). 

In the non-perturbative regime of the Anderson model, the physical quantities such as 
the magnetization $m$ must be universal functions of $T/T_K$ or $h/T_K$. 
This means that $\tau$ should be related to the characteristic temperature $T_K$ 
of the orbital Kondo effect yielded by the inversion symmetry of ISIM. The relation 
between $\tau$ and $T_K$ can be obtained using Friedel sum rule and analytical results 
for the magnetization $m(h)$ of the Anderson model at zero temperature. 

\section{Impurity occupation numbers and scattering phase shifts}
\label{Friedel Sum Rule}
\begin{figure}
\centerline{
\includegraphics[width=\columnwidth]{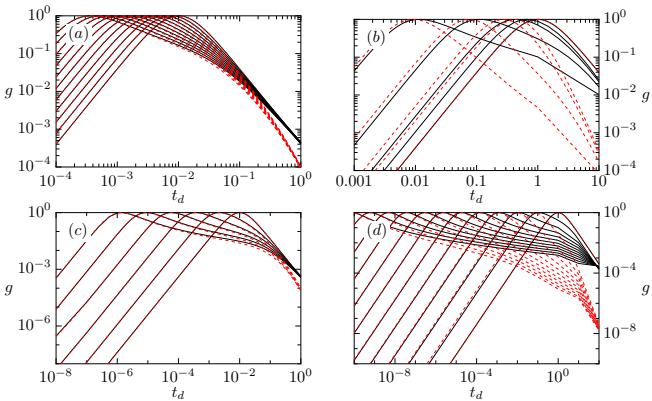}
}
\caption{(Color online) Values of $g$ obtained from the NRG spectrum 
(black solid lines, Eq.~\eqref{transmission}) and approximate values 
$\tilde g$ (red dashed lines, Eq.~\eqref{FSR}), as a function of $t_d$. 
Increasing $U$ moves the conductance peak towards the left side. Figure (a): 
$t_c=0.1$ and $U=0, 0.01, 0.02, \ldots, 1$. Figure (b): $t_c=1$ and $U=0,1,2,5,10$. 
Figure (c): $t_c=0.1$ and $U=0, 0.05, 0.1, 0.15,0.2, 0.25$. Figure (d): 
$t_c=1$ and $U=0,5,10,15,\ldots, 55$. 
\label{FIGURE16} 
}
\end{figure}

For having $g$, another method consists in using NRG for calculating the average 
impurity occupation numbers $\langle n_{\mathrm e/\mathrm o}\rangle $ of the even and odd orbitals of the nanosystem. 
The difference between the scattering phase shifts can be given in terms of $\langle n_{\mathrm e/\mathrm o}\rangle$, 
if one assumes an approximate form of the Friedel sum rule (FSR): 
\begin{equation}
\delta_{\mathrm e}-\delta_{\mathrm o} \approx \pi \left(\left\langle n_{\mathrm e}\right\rangle - 
\left\langle n_{\mathrm o}\right\rangle \right)\,,  
\label{FSR}
\end{equation}
and one gets an approximate value $\tilde g$ for $g$ from this estimate of 
$\delta_{\mathrm e}-\delta_{\mathrm o}$ using Eq.~\eqref{phaseshift}:
\begin{equation}
\tilde g =\sin^2(\pi \left(\left\langle n_{\mathrm e}\right\rangle-\left\langle n_{\mathrm o}\right\rangle \right)\,.
\label{gapprox}
\end{equation}

This approximate FSR is often used (see for instance Ref.~\onlinecite{ng-lee}) and 
allows us to obtain $g$ from the zero temperature impurity magnetization $m(h)$ 
of the Anderson model with magnetic field $h$. This is particularly interesting 
since $m(h)$ is a physical quantity for which exact results have been obtained 
with the Bethe-Ansatz by Tsvelick and Wiegmann. Unfortunately, Eq~\eqref{FSR} is 
only an approximation, and not the true FSR, as pointed out by Simon and Affleck in 
a study \cite{simon} of persistent currents through a quantum dot at Kondo resonance. 
The generalization of FSR by Langreth \cite{langreth,hewson} shows that the phase shifts 
are proportional to the number of electrons displaced by the impurity, ``among which are 
included not only the d electrons, but also some of the conduction electrons''. For ISIM, 
this means that the displaced electrons are not only those inside the interacting region, 
but displaced electrons in the neighboring parts of the leads have to be included too for 
obtaining the phase shifts from the occupation numbers via FSR. At first sight, one can 
expect that Eq.~\eqref{FSR} could be used only if the scattering region is weakly coupled 
to the attached leads. Our results show that this is less simple. 

  The difference between the values of $g$ obtained directly from the NRG spectra 
[Eq.~\eqref{transmission}] and the approximated values $\tilde g$ are given as a function of 
$t_d$ for weak ($t_c=0.1$, Fig.~\ref{FIGURE16}a and d) and large values ($t_c=1$, 
Fig.~\ref{FIGURE16}b and c) of the coupling, and for weak (Fig.~\ref{FIGURE16}a and c) 
or large values (Fig.~\ref{FIGURE16}b and d) of $U$. Even for a weak coupling $t_c=0.1$, 
where one could expect a negligible displaced charge outside the nanosystem,  
$\tilde g=g$ only when ISIM is near the SC fixed point ($t_d \leq \tau$). When $t_d > \tau$, 
$\tilde g \neq g$. For a larger coupling ($t_c=1$), one can notice also differences between 
$g$ and $\tilde g$ even when $t_d \leq \tau$ in the perturbative regime ($U<\pi t_c^2$). 
One concludes that $g$ can be obtained from $\tilde g$ with a good accuracy 
only in the non-perturbative regime where ISIM exhibits an orbital Kondo effect ($U > 
t_c^2/A$ and $t_d \leq \tau$). Otherwise, the difference $\delta_{\mathrm e}-\delta_{\mathrm o}$ is not 
given by the difference $\left\langle n_{\mathrm e}\right\rangle -\left\langle n_{\mathrm o}\right\rangle $ evaluated inside the 
nanosystem, but depends also on the occupation numbers outside the nanosystem. 

The validity of the approximate FSR in the non-perturbative regime when $t_d < \tau$ can be 
explained by the following argument: the conduction electrons which are displaced to screen 
the impurity pseudo-spin (forming a singlet state with the impurity) are only a negligible 
fraction $\approx T_K/E_F$ of the conduction electrons. To neglect this fraction induces an 
error $\propto \tau / t_h$ which cannot be seen in the curves shown in Fig.~\ref{FIGURE16} 
when $\tau$ is very small (see Fig.~\ref{FIGURE11}). When the system is not in the orbital 
Kondo regime ($U < \pi t_c^2$ or $t_d > \tau$), the number of displaced electrons becomes 
much larger, and the difference between $g$ and $\tilde g$ can be seen in Fig.~\ref{FIGURE16}. 
 
\section{Zero temperature magnetization of the symmetric Anderson model}
\label{Bethe-Ansatz}

  The exact solution of the Anderson model can be obtained using Bethe-Ansatz. 
Tsvelick and Wiegmann have solved the Bethe-Ansatz equations for the Anderson model in an 
arbitrary magnetic field. Let us summarize their results \cite{tsvelick-wiegmann1,tsvelick-wiegmann2} 
for the symmetric case, which were obtained assuming the continuum limit and an infinite bandwidth 
($t_h \to \infty$) for an Anderson impurity coupled to a 3D bath of non-interacting electrons. 
Since the bath for ISIM is provided by non-interacting electrons free to move on a semi-infinite 
1D tight-binding lattice, and since we give results for values of $t_c$, $t_d$ and $U$ which are 
not always small compared to $t_h$, one cannot rule out certain quantitative differences between 
the results of Refs.~\onlinecite{tsvelick-wiegmann1,tsvelick-wiegmann2} and our numerical results. This 
may concern the numerical prefactors in the expression of $T_K$ or the constants in the universal 
functions describing the magnetization $m(h)$. However, a qualitative agreement should be expected. 
The Kondo temperature of the Anderson model reads
\begin{equation}
T_K=  F \sqrt{Ut_c^2} \exp -\left( \frac{\pi U}{8 t_c^2} \right)\,,
\label{Kondo}
\end{equation}
where $F$ is a prefactor which depends on the definition of $T_K$ (which varies \cite{tsvelick-wiegmann1} 
from one author to another) and is modified \cite{haldane} if the bandwidth of the bath of conduction 
electrons is finite or infinite. $F=\sqrt{2}/\pi$ for the infinite bandwidth Anderson 
model \cite{tsvelick-wiegmann1,tsvelick-wiegmann2}, while $F=0.364 \sqrt{2/\pi}$ if the bandwidth is taken 
finite \cite{haldane}.

 When $U \geq \Gamma$  (non-perturbative regime), the Bethe-Ansatz results 
for the impurity magnetization $m(h)$ can be divided in 
three characteristic regimes as the magnetic field $h$ increases at $T=0$. 

A SC regime for low fields ($h < T_K$) where the magnetization is given by 
a universal function of $h/T_K$:  
\begin{equation}
m(h)= \frac{h}{2\pi T_K}\,, 
\end{equation}
followed by a LM regime for intermediate fields ($T_k < h < \sqrt{U\Gamma}$) 
where $m(h)$ is given by another universal function of $h/T_K$ which can be expended as 
\begin{equation}
m(h)\approx \frac{1}{2} \left( 1-\frac{1}{\ln(\frac{h}{T_K})} +\ldots \right)\,, 
\label{m-lm}
\end{equation}
before having a PO regime for strong fields ($ h > \sqrt{U\Gamma}$) 
(denoted FO regime in Ref.~\onlinecite{tsvelick-wiegmann1}) where 
\begin{equation}
m(h)\approx \frac{1}{2} \left(1-\frac{2\Gamma}{\pi h} + \ldots \right) 
\label{m-po}
\end{equation}
becomes independent of the interaction $U$.

 When $U \approx \Gamma$, there is direct transition from a non-magnetic regime 
where $m(h)\approx h/\Gamma$ towards the free orbital regime where the behavior of $m(h)$ 
is given by Eq.~\eqref{m-po}.
 
\section{Characteristic Energy Scale and Universal Scaling Functions}
\label{non-perturbative} 

In the perturbative regime ($U < \pi t_c^2$), the conductance is well described by HF theory, 
which yields $g=4/(v/t_c^2+t_c^2/v)^2$ with $v$ given by Eq.~\eqref{hopping-HF}. The scale $\tau$ 
takes a value $\tau_{HF}$ given by Eq.~\eqref{scale-HF}. 

In the non-perturbative regime ($U > \pi t_c^2$), let us revisit our numerical results for $g$ using 
the exact expressions giving the magnetization $m(h)$ for an Anderson model (3D bath of conduction 
electrons, wide band limit where $t_h \to \infty$) which is not exactly the Anderson model corresponding 
to ISIM. 

The conductance $\tilde g(t_d)$ reads 
\begin{equation}
\tilde g(t_d)= \sin^2 \left(2\pi m(t_d) \right) 
\end{equation}
where $m(t_d)$ is the pseudo-magnetization of the nanosystem with pseudo-spin
$1/2$ and reads
\begin{equation}
m(t_d)= \frac{\left\langle n_{\mathrm e}(t_d)\right\rangle -\left\langle
n_{\mathrm o}(t_d)\right\rangle }{2}\,. 
\end{equation}
In the SC limit ($t_d<T_K$) where $g$ and $\tilde g$ coincide, this gives
\begin{equation}
\tilde g(t_d)= \sin^2(t_d/T_k) \approx \left(\frac{t_d}{T_K} \right)^2 \approx g(t_d)\,,  
\end{equation}
while the conductance $g$ extracted from the NRG spectra reads 
\begin{equation}
g(t_d)=\left(\frac{2}{t_d/\tau+\tau/t_d}\right)^2 \approx 4 \left(\frac{t_d}{\tau}\right)^2\,. 
\end{equation}
This yields a relation between the characteristic scale $\tau$ of ISIM and the Kondo 
temperature $T_K$ of the Anderson model: 
\begin{equation}
\tau(U,t_c^2)=2 T_K(U,t_c^2)\,.
\label{Kondo-tau-1}
\end{equation}

\begin{figure}
\centerline{
\includegraphics[width=\columnwidth]{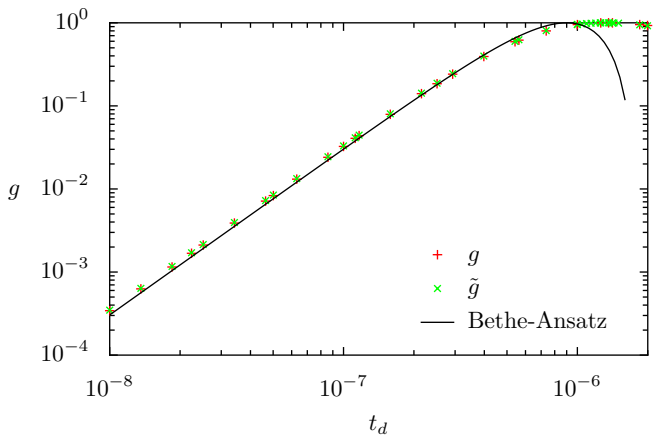}
}
\caption{(Color online) SC regime: Values of $g$ (red plus), $\tilde g$ 
(green cross) obtained from the NRG algorithm and Bethe-Ansatz expression 
$\sin^2(2t_d/\tau)$ (solid line) as a function of $t_d$ for $U=0.25$ and $t_c=0.1$. 
The value of $\tau$ used in the Bethe-Ansatz expression have been obtained from 
the relation $\tau=2T_K$, with $T_K$ given by Eq.~\eqref{Kondo-tau-2}.
}  
\label{FIGURE17} 
\end{figure}

Using for $T_K$ Eq.~\eqref{Kondo} with the finite bandwidth prefactor $F=0.364 \sqrt{2/\pi}$ 
given in Ref.~\onlinecite{haldane}, one gets for $\tau$ the analytical expression
\begin{equation}
\tau(U,t_c)=0.728 \sqrt{\frac{2t_c^2U}{\pi}} \exp -\left(\frac{\pi U}{8 t_c^2}\right)\,.
\label{Kondo-tau-2}
\end{equation}
Eq.~\eqref{Kondo-tau-2} describes very well the numerical values of $\tau(U,t_c)$ obtained 
from the NRG spectra and the condition $g(t_d=\tau)=1$ in the non-perturbative regime, 
as shown in Fig.~\ref{FIGURE11} for $U > t_c^2/A$. Moreover, the behavior of $g$ calculated 
for $U=0.25$ and $t_c=0.1$ is given in Fig.~\ref{FIGURE17} as a function of $t_d$ when 
$t_d < \tau$. One can see that the expression $\tilde g =\sin^2(2 t_d/\tau)$ with the value 
of $\tau$ calculated from Eq.~\eqref{Kondo-tau-2} ($\tau=1.5825 10^{-6}$) describes the 
behavior of $g$ or $\tilde g$ calculated using the NRG algorithm when $t_d < \tau$.
\begin{figure}
\centerline{
\includegraphics[width=\columnwidth]{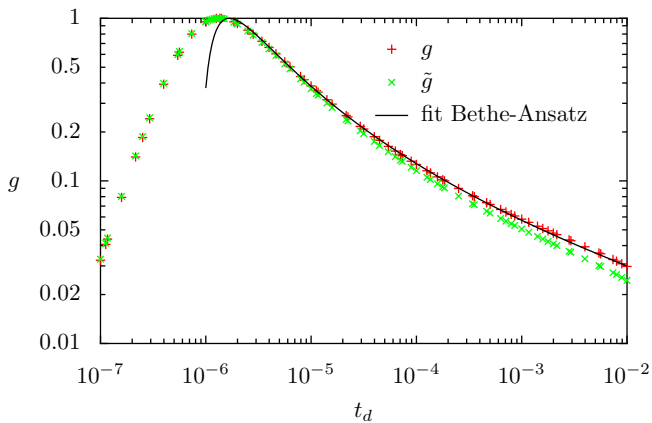}
}
\caption{(Color online) 
LM regime:  NRG values of $g$ (red plus) and $\tilde g$ (green cross) as a function of $t_d$ 
for $U=0.25$ and $t_c=0.1$. The solid line (fit Bethe-Ansatz) corresponds to Eq.~\eqref{g-lm-2} 
with the value of $\tau=2T_K$ given by Eq.~\eqref{Kondo-tau-2} and used in Fig.~\ref{FIGURE17}.}  
\label{FIGURE18} 
\end{figure}
 
Fig.~\ref{FIGURE18} gives the NRG values of $g$ and $\tilde g$ as a function of $t_d$ for 
$U=0.25$ and $t_c=0.1$ in the LM regime. One can see that $g \approx \tilde g$ around the 
transmission peak, but becomes slightly different when $t_d > \tau$. Using Eq.~\eqref{m-lm} 
for $m(t_d)$, the NRG values of $\tilde g(t_d)$ shown in Fig.~\ref{FIGURE18} are not reproduced 
by $\sin^2\left(2\pi m(t_d)\right)$. We explain this failure by the fact that the Anderson model corresponding to 
Eq.~\eqref{m-lm} is not exactly the Anderson model corresponding to ISIM. This might give different 
constants in the function given in Eq.~\eqref{m-lm}. However, we have been able to find a function 
of $X=t_d/\tau$ inspired by the form of $m(h/T_K)$ given in Eq.~\eqref{m-lm} and which fits very 
well the values of $g$ in the LM regime: 
\begin{equation}
g(X)= \sin^2 \left(\pi(2.0175-\frac{0.7388}{\ln (2.8573 X)})\right)\,. 
\label{g-lm-2}
\end{equation}
 As shown in Fig.~\ref{FIGURE18}, such a fit with the value of $\tau=2T_K$ used in Fig.~\ref{FIGURE17} 
allows us to describe $g(t_d)$. Eq.~\eqref{g-lm-2} gives an excellent approximation of the universal curve 
of $g(t_d/\tau)$ in the LM regime.

When $t_d > t_c \sqrt{U}$ (PO regime), $g$ can be described by the non-interacting expression 
$4(t_d/t_c^2+t_c^2/t_d)^{-2}$, which achieves the complete description of $g(t_d,U,t_C)$ 
in the symmetric case by analytical expressions.

\section{Summary and Perspective}
\label{Conclusion}
 
 When $t_d \le \tau$, we have shown that the quantum conductance is given by 
a universal function of the ratio $X=t_d/\tau$ of two energies. This universal 
function $g(X)=4(X+X^{-1})^{-2}$ characterizes the non-interacting limit, where 
the isolated nanosystem has two levels of energy $V_G \pm t_d$ with a level spacing 
$\Delta=2t_d$. Those levels have a width $\Gamma=t_c^2$ when the nanosystem is coupled to 
leads at $E_F=0$. Therefore the ratio $X$ is also the ratio $\Delta / 2 \Gamma$. We have 
found that $g(X)$ remains unchanged when the electrons interact inside the nanosystem, if 
one adds a term $\propto U$ to the  broadening $\Gamma$ in the perturbative regime. When 
$U$ becomes larger, there is a non-perturbative regime where a more complicated 
many-body resonance appears at $E_F$. In that case, the relation $\tau=2 T_K$ which we have 
obtained is consistent with the fact that the width $\Gamma$ of the nanosystem levels 
becomes the Kondo temperature $T_K$. We have thus shown that the interaction $U$ leaves the 
function $g(X)$ unchanged in the SC regime, renormalizing only the level broadening $\Gamma$. 

To have a conductance which is given by the ratio of two characteristic energies 
and which stays at zero temperature on a universal curve when this ratio varies 
is reminiscent of the scaling theory of localization \cite{abrahams}. Recently, it 
has been found \cite{fleury-waintal} using numerical Quantum Monte Carlo simulations that the 
$\beta(g)$-function characterizes not only the non-interacting limit \cite{abrahams}, but 
also 2D disordered systems with Coulomb interactions. We have given another example of a 
universal function which remains unchanged when the electrons interact, the interaction 
renormalizing only the characteristic scale ($\Gamma$ for the dimensionless ratio $\Delta/\Gamma$ 
in this study, the localization length $\xi$ for the dimensionless ratio $L/\xi$ in 2D disordered 
systems \cite{fleury-waintal}). 

 When $\tau <t_d < t_c\sqrt{U}$, a local moment is formed in the equivalent Anderson model
and the universal function $g(X)$ is not given by the non-interacting limit. Adapting an 
analytical expression describing the magnetization of the Anderson model, we have proposed 
an analytical form which reproduces the universal function $g(X)$ in the LM regime.

 We have shown that the interacting region becomes perfectly transparent when $t_d=\tau$. 
In the non-perturbative regime, this corresponds to a nanosystem level spacing $\Delta \approx T_K$. 
This result shown using ISIM is a particular illustration of the minimal realization 
of the orbital Kondo effect in a quantum dot with two leads, which has been studied 
by Silvestrov and Imry~\cite{silvestrov} for more general setups. The role of the conduction 
electron spin is played by the lead index in Ref.~\onlinecite{silvestrov}, while it is played by 
the even and odd orbitals for ISIM. In the two cases, the Kondo effect takes place if there 
are two close levels in a dot populated by a single electron, and the conductance at $T=0$ is 
zero at the SU(2) symmetric point ($t_d=0$ for ISIM), while it reaches the unitary limit 
$G=e^2/h$ for some finite value $\approx T_K$ of the level splitting $\Delta$. However, 
the prediction made in Ref.~\onlinecite{silvestrov}, that for temperature $T>T_K$ the conductance 
becomes maximal if the levels are exactly degenerate, cannot be valid for ISIM where level 
degeneracy means no coupling between the left and right leads ($t_d=0$). 

We have shown that the quantum conductance of an interacting nanosystem coupled to non-interacting 
1D leads can be described with a universal function $g(X)$. This concept, with a similar function 
$g(X)$, must remain valid if one couples the nanosystem to non-interacting 2D or 3D leads instead 
of strictly 1D leads. This can be understood if one considers the Anderson models of pseudo-spin $1/2$ 
particles corresponding to ISIM with leads of dimension D. The Kondo physics of such models, where 
the quantum impurity is coupled to a bath of dimension D, is qualitatively independent of the used bath. 
A change of the dimension of the leads modifies only the dependence of the nanosystem level width $\Gamma$ 
upon $t_c$ ($\Gamma \propto t_c^2$ in all dimensions D, but with factors which depend on D), and hence the 
dependence of the Kondo temperature upon $t_c$. This makes likely that the universal aspects of $g$ 
obtained in a pure 1D limit using ISIM do characterize also more general spinless models, where the 
nanosystem and the leads would be created in gated 2D semiconductor heterostructures. In that case, 
ISIM is a simplified model which could describe quantum transport of spin polarized electrons through 
an inversion-symmetric double-dot setup, as a function of the inter-dot coupling. Since such a coupling 
can be easily varied if the two dots are 
coupled by a quantum point contact, it will be interesting to check whether the quantum conductance 
of such setup is given by a universal function of the dimensionless inter-dot coupling $t_d/T_K$ when 
$T \to 0$. For observing the orbital Kondo regime using such a setup, a large capacitive inter-dot 
coupling will be necessary.  It will be also interesting to introduce the spin 1/2 of electrons in 
ISIM for studying the role of $t_d$ upon the SU(4)-Kondo effect, as we have studied its role upon the 
SU(2)-Kondo effect using spinless fermions. The possibility of observing SU(4)-symmetric Fermi liquid 
state in a symmetric double quantum dot system with strong capacitive inter-dot coupling has been 
discussed in Ref.~\onlinecite{borda}.

Eventually, this study was restricted to the symmetric case, leaving to a following work the study of the 
asymmetric case ($E_F \neq 0$, $V_G \neq -U/2$), where the role of $t_d$ upon the valence-fluctuation 
fixed point remains to be investigated.
 
We thank Denis Ullmo for very useful discussions and the RTRA ``Triangle de la Physique'' of 
Palaiseau-Orsay-Saclay for financial support.


\begin{thebibliography}{36}
\expandafter\ifx\csname natexlab\endcsname\relax\def\natexlab#1{#1}\fi
\expandafter\ifx\csname bibnamefont\endcsname\relax
  \def\bibnamefont#1{#1}\fi
\expandafter\ifx\csname bibfnamefont\endcsname\relax
  \def\bibfnamefont#1{#1}\fi
\expandafter\ifx\csname citenamefont\endcsname\relax
  \def\citenamefont#1{#1}\fi
\expandafter\ifx\csname url\endcsname\relax
  \def\url#1{\texttt{#1}}\fi
\expandafter\ifx\csname urlprefix\endcsname\relax\def\urlprefix{URL }\fi
\providecommand{\bibinfo}[2]{#2}
\providecommand{\eprint}[2][]{\url{#2}}

\bibitem[{\citenamefont{Meir and Wingreen}(1992)}]{meir-wingreen}
\bibinfo{author}{\bibfnamefont{Y.}~\bibnamefont{Meir}} \bibnamefont{and}
  \bibinfo{author}{\bibfnamefont{N.~S.} \bibnamefont{Wingreen}},
  \bibinfo{journal}{Phys. Rev. Lett.} \textbf{\bibinfo{volume}{68}},
  \bibinfo{pages}{2512} (\bibinfo{year}{1992}).

\bibitem[{\citenamefont{Molina et~al.}(2004)\citenamefont{Molina,
  Schmitteckert, Weinmann, Jalabert, , Ingold, and Pichard}}]{molina1}
\bibinfo{author}{\bibfnamefont{R.~A.} \bibnamefont{Molina}},
  \bibinfo{author}{\bibfnamefont{P.}~\bibnamefont{Schmitteckert}},
  \bibinfo{author}{\bibfnamefont{D.}~\bibnamefont{Weinmann}},
  \bibinfo{author}{\bibfnamefont{R.~A.} \bibnamefont{Jalabert}}, ,
  \bibinfo{author}{\bibfnamefont{G.-L.} \bibnamefont{Ingold}},
  \bibnamefont{and} \bibinfo{author}{\bibfnamefont{J.-L.}
  \bibnamefont{Pichard}}, \bibinfo{journal}{Eur. Phys. J. B}
  \textbf{\bibinfo{volume}{39}}, \bibinfo{pages}{107} (\bibinfo{year}{2004}).

\bibitem[{\citenamefont{Molina et~al.}(2005)\citenamefont{Molina, Weinmann, and
  Pichard}}]{molina2}
\bibinfo{author}{\bibfnamefont{R.~A.} \bibnamefont{Molina}},
  \bibinfo{author}{\bibfnamefont{D.}~\bibnamefont{Weinmann}}, \bibnamefont{and}
  \bibinfo{author}{\bibfnamefont{J.-L.} \bibnamefont{Pichard}},
  \bibinfo{journal}{Eur. Phys. J. B} \textbf{\bibinfo{volume}{48}},
  \bibinfo{pages}{243} (\bibinfo{year}{2005}).

\bibitem[{\citenamefont{Weinmann et~al.}(2008)\citenamefont{Weinmann, Jalabert,
  Freyn, Ingold, and Pichard}}]{weinmann}
\bibinfo{author}{\bibfnamefont{D.}~\bibnamefont{Weinmann}},
  \bibinfo{author}{\bibfnamefont{R.~A.} \bibnamefont{Jalabert}},
  \bibinfo{author}{\bibfnamefont{A.}~\bibnamefont{Freyn}},
  \bibinfo{author}{\bibfnamefont{G.-L.} \bibnamefont{Ingold}},
  \bibnamefont{and} \bibinfo{author}{\bibfnamefont{J.-L.}
  \bibnamefont{Pichard}}, \bibinfo{journal}{Eur. Phys. J. B}
  \textbf{\bibinfo{volume}{66}}, \bibinfo{pages}{239} (\bibinfo{year}{2008}).

\bibitem[{\citenamefont{Asada et~al.}(2006)\citenamefont{Asada, Freyn, and
  Pichard}}]{asada}
\bibinfo{author}{\bibfnamefont{Y.}~\bibnamefont{Asada}},
  \bibinfo{author}{\bibfnamefont{A.}~\bibnamefont{Freyn}}, \bibnamefont{and}
  \bibinfo{author}{\bibfnamefont{J.-L.} \bibnamefont{Pichard}},
  \bibinfo{journal}{Eur. Phys. J. B} \textbf{\bibinfo{volume}{53}},
  \bibinfo{pages}{109} (\bibinfo{year}{2006}).

\bibitem[{\citenamefont{Freyn and Pichard}(2006)}]{freyn-pichard1}
\bibinfo{author}{\bibfnamefont{A.}~\bibnamefont{Freyn}} \bibnamefont{and}
  \bibinfo{author}{\bibfnamefont{J.-L.} \bibnamefont{Pichard}},
  \bibinfo{journal}{Phys. Rev. Lett.} \textbf{\bibinfo{volume}{98}},
  \bibinfo{pages}{186401} (\bibinfo{year}{2007}).

\bibitem[{\citenamefont{Freyn and Pichard}(2007)}]{freyn-pichard2}
\bibinfo{author}{\bibfnamefont{A.}~\bibnamefont{Freyn}} \bibnamefont{and}
  \bibinfo{author}{\bibfnamefont{J.-L.} \bibnamefont{Pichard}},
  \bibinfo{journal}{Eur. Phys. J. B} \textbf{\bibinfo{volume}{58}},
  \bibinfo{pages}{279} (\bibinfo{year}{2007}).

\bibitem[{\citenamefont{Freyn and Pichard}(2008)}]{freyn-pichard3}
\bibinfo{author}{\bibfnamefont{A.}~\bibnamefont{Freyn}},
  \bibinfo{author}{\bibfnamefont{I.} \bibnamefont{Kleftogiannis}} \bibnamefont{and}
  \bibinfo{author}{\bibfnamefont{J.-L.} \bibnamefont{Pichard}},
  \bibinfo{journal}{Phys. Rev. Lett.} \textbf{\bibinfo{volume}{100}},
  \bibinfo{pages}{226802} (\bibinfo{year}{2008}).

\bibitem[{\citenamefont{Krishna-murthy
  et~al.}(1975)\citenamefont{Krishna-murthy, Wilson, and
  Wilkins}}]{krishna-murthy1}
\bibinfo{author}{\bibfnamefont{H.~R.} \bibnamefont{Krishna-murthy}},
  \bibinfo{author}{\bibfnamefont{K.~G.} \bibnamefont{Wilson}},
  \bibnamefont{and} \bibinfo{author}{\bibfnamefont{J.~W.}
  \bibnamefont{Wilkins}}, \bibinfo{journal}{Phys. Rev. Lett.}
  \textbf{\bibinfo{volume}{35}}, \bibinfo{pages}{1101} (\bibinfo{year}{1975}).

\bibitem[{\citenamefont{Krishna-murthy
  et~al.}(1980{\natexlab{a}})\citenamefont{Krishna-murthy, Wilkins, and
  Wilson}}]{krishna-murthy2}
\bibinfo{author}{\bibfnamefont{H.~R.} \bibnamefont{Krishna-murthy}},
  \bibinfo{author}{\bibfnamefont{J.~W.} \bibnamefont{Wilkins}},
  \bibnamefont{and} \bibinfo{author}{\bibfnamefont{K.~G.}
  \bibnamefont{Wilson}}, \bibinfo{journal}{Phys. Rev. B}
  \textbf{\bibinfo{volume}{21}}, \bibinfo{pages}{1003}
  (\bibinfo{year}{1980}{\natexlab{a}}).

\bibitem[{\citenamefont{Krishna-murthy
  et~al.}(1980{\natexlab{b}})\citenamefont{Krishna-murthy, Wilkins, and
  Wilson}}]{krishna-murthy3}
\bibinfo{author}{\bibfnamefont{H.~R.} \bibnamefont{Krishna-murthy}},
  \bibinfo{author}{\bibfnamefont{J.~W.} \bibnamefont{Wilkins}},
  \bibnamefont{and} \bibinfo{author}{\bibfnamefont{K.~G.}
  \bibnamefont{Wilson}}, \bibinfo{journal}{Phys. Rev. B}
  \textbf{\bibinfo{volume}{21}}, \bibinfo{pages}{1044}
  (\bibinfo{year}{1980}{\natexlab{b}}).

\bibitem[{\citenamefont{Bulla et~al.}(2008)\citenamefont{Bulla, Costi, and
  Pruschke}}]{bulla}
\bibinfo{author}{\bibfnamefont{R.}~\bibnamefont{Bulla}},
  \bibinfo{author}{\bibfnamefont{T.~A.} \bibnamefont{Costi}}, \bibnamefont{and}
  \bibinfo{author}{\bibfnamefont{T.}~\bibnamefont{Pruschke}},
  \bibinfo{journal}{Rev. Mod. Phys.} \textbf{\bibinfo{volume}{80}},
  \bibinfo{pages}{395} (\bibinfo{year}{2008}).

\bibitem[{\citenamefont{Hewson}(1993{\natexlab{a}})}]{hewson}
\bibinfo{author}{\bibfnamefont{A.~C.} \bibnamefont{Hewson}},
  \emph{\bibinfo{title}{The Kondo Problem To Heavy Fermions}}
  (\bibinfo{publisher}{Cambridge University Press},
  \bibinfo{year}{1993}{\natexlab{a}}).

\bibitem[{\citenamefont{Tsvelick and Wiegmann}(1983)}]{tsvelick-wiegmann1}
\bibinfo{author}{\bibfnamefont{A.~M.} \bibnamefont{Tsvelick}} \bibnamefont{and}
  \bibinfo{author}{\bibfnamefont{P.~B.} \bibnamefont{Wiegmann}},
  \bibinfo{journal}{Advances in Phys.} \textbf{\bibinfo{volume}{32}},
  \bibinfo{pages}{453} (\bibinfo{year}{1983}).

\bibitem[{\citenamefont{Wiegmann and Tsvelick}(1983)}]{tsvelick-wiegmann2}
\bibinfo{author}{\bibfnamefont{P.~B.} \bibnamefont{Wiegmann}} \bibnamefont{and}
  \bibinfo{author}{\bibfnamefont{A.~M.} \bibnamefont{Tsvelick}},
  \bibinfo{journal}{J. Phys. C: Solid State Phys.}
  \textbf{\bibinfo{volume}{16}}, \bibinfo{pages}{2281} (\bibinfo{year}{1983}).

\bibitem[{\citenamefont{Goldhaber-Gordon
  et~al.}(1998)\citenamefont{Goldhaber-Gordon, Shtrikman, Mahalu,
  Abusch-Magder, Meirav, and Kastner}}]{goldhaber-gordon}
\bibinfo{author}{\bibfnamefont{D.}~\bibnamefont{Goldhaber-Gordon}},
  \bibinfo{author}{\bibfnamefont{H.}~\bibnamefont{Shtrikman}},
  \bibinfo{author}{\bibfnamefont{D.}~\bibnamefont{Mahalu}},
  \bibinfo{author}{\bibfnamefont{D.}~\bibnamefont{Abusch-Magder}},
  \bibinfo{author}{\bibfnamefont{U.}~\bibnamefont{Meirav}}, \bibnamefont{and}
  \bibinfo{author}{\bibfnamefont{M.~A.} \bibnamefont{Kastner}},
  \bibinfo{journal}{Nature (London)} \textbf{\bibinfo{volume}{391}},
  \bibinfo{pages}{156} (\bibinfo{year}{1998}).

\bibitem[{\citenamefont{Cronenwett et~al.}(1998)\citenamefont{Cronenwett,
  Oosterkamp, and Kouvanhoven}}]{cronenwett}
\bibinfo{author}{\bibfnamefont{S.~M.} \bibnamefont{Cronenwett}},
  \bibinfo{author}{\bibfnamefont{T.~H.} \bibnamefont{Oosterkamp}},
  \bibnamefont{and} \bibinfo{author}{\bibfnamefont{L.~P.}
  \bibnamefont{Kouvanhoven}}, \bibinfo{journal}{Science}
  \textbf{\bibinfo{volume}{281}}, \bibinfo{pages}{540} (\bibinfo{year}{1998}).

\bibitem[{\citenamefont{Grobis et~al.}(2008)\citenamefont{Grobis, Rau, Potok,
  Shtrikman, and Goldhaber-Gordon}}]{grobis}
\bibinfo{author}{\bibfnamefont{M.}~\bibnamefont{Grobis}},
  \bibinfo{author}{\bibfnamefont{I.~G.} \bibnamefont{Rau}},
  \bibinfo{author}{\bibfnamefont{R.~M.} \bibnamefont{Potok}},
  \bibinfo{author}{\bibfnamefont{H.}~\bibnamefont{Shtrikman}},
  \bibnamefont{and}
  \bibinfo{author}{\bibfnamefont{D.}~\bibnamefont{Goldhaber-Gordon}},
  \bibinfo{journal}{Phys. Rev. Lett.} \textbf{\bibinfo{volume}{100}},
  \bibinfo{pages}{246601} (\bibinfo{year}{2008}).

\bibitem[{\citenamefont{Delattre et~al.}(2009)\citenamefont{Delattre,
  Feuillet-Palma, Herrmann, Morfin, Berroir, F\`eve, Placais, Glattli, Choi,
  Mora, and Kontos}}]{delattre}
\bibinfo{author}{\bibfnamefont{T.}~\bibnamefont{Delattre}},
  \bibinfo{author}{\bibfnamefont{C.}~\bibnamefont{Feuillet-Palma}},
  \bibinfo{author}{\bibfnamefont{L.~G.} \bibnamefont{Herrmann}},
  \bibinfo{author}{\bibfnamefont{P.}~\bibnamefont{Morfin}},
  \bibinfo{author}{\bibfnamefont{J.-M.} \bibnamefont{Berroir}},
  \bibinfo{author}{\bibfnamefont{G.}~\bibnamefont{F\`eve}},
  \bibinfo{author}{\bibfnamefont{B.}~\bibnamefont{Placais}},
  \bibinfo{author}{\bibfnamefont{D.~C.} \bibnamefont{Glattli}},
  \bibinfo{author}{\bibfnamefont{M.-S.} \bibnamefont{Choi}},
  \bibinfo{author}{\bibfnamefont{C.}~\bibnamefont{Mora}}, 
  \bibnamefont{and}
  \bibinfo{author}{\bibfnamefont{T.}~\bibnamefont{Kontos}}, 
  \bibinfo{journal}{Nature Physics} \textbf{\bibinfo{volume}{5}},
  \bibinfo{pages}{208} (\bibinfo{year}{2009}).

\bibitem[{\citenamefont{Kaul et~al.}(2006)\citenamefont{Kaul, Zar\'and,
  Chandrasekharan, Ullmo, and Baranger}}]{kaul}
\bibinfo{author}{\bibfnamefont{R.~K.}~\bibnamefont{Kaul}},
  \bibinfo{author}{\bibfnamefont{G.}~\bibnamefont{Zar\'and}},
  \bibinfo{author}{\bibfnamefont{S.}~\bibnamefont{Chandrasekharan}},
  \bibinfo{author}{\bibfnamefont{D.}~\bibnamefont{Ullmo}}, \bibnamefont{and}
  \bibinfo{author}{\bibfnamefont{H.~U.}~\bibnamefont{Baranger}},
  \bibinfo{journal}{Phys. Rev. Lett.} \textbf{\bibinfo{volume}{96}},
  \bibinfo{pages}{176802} (\bibinfo{year}{2006}).

\bibitem[{\citenamefont{Mehta and Andrei}(2006)}]{mehta}
\bibinfo{author}{\bibfnamefont{P.}~\bibnamefont{Mehta}} \bibnamefont{and}
  \bibinfo{author}{\bibfnamefont{N.}~\bibnamefont{Andrei}},
  \bibinfo{journal}{Phys. Rev. Lett.} \textbf{\bibinfo{volume}{96}},
  \bibinfo{pages}{216802} (\bibinfo{year}{2006}).

\bibitem[{\citenamefont{Boulat et~al.}(2008)\citenamefont{Boulat, Saleur, and
  Schmitteckert}}]{boulat}
\bibinfo{author}{\bibfnamefont{E.}~\bibnamefont{Boulat}},
  \bibinfo{author}{\bibfnamefont{H.}~\bibnamefont{Saleur}}, \bibnamefont{and}
  \bibinfo{author}{\bibfnamefont{P.}~\bibnamefont{Schmitteckert}},
  \bibinfo{journal}{Phys. Rev. Lett.} \textbf{\bibinfo{volume}{101}},
  \bibinfo{pages}{140601} (\bibinfo{year}{2008}).

\bibitem[{\citenamefont{Dhar et~al.}(2008)\citenamefont{Dhar, Sen, and
  Roy}}]{dhar}
\bibinfo{author}{\bibfnamefont{A.}~\bibnamefont{Dhar}},
  \bibinfo{author}{\bibfnamefont{D.}~\bibnamefont{Sen}}, \bibnamefont{and}
  \bibinfo{author}{\bibfnamefont{D.}~\bibnamefont{Roy}},
  \bibinfo{journal}{Phys. Rev. Lett.} \textbf{\bibinfo{volume}{101}},
  \bibinfo{pages}{066805} (\bibinfo{year}{2008}).

\bibitem[{\citenamefont{Anderson}(1961)}]{anderson}
\bibinfo{author}{\bibfnamefont{P.~W.} \bibnamefont{Anderson}},
  \bibinfo{journal}{Phys. Rev. B} \textbf{\bibinfo{volume}{124}},
  \bibinfo{pages}{41} (\bibinfo{year}{1961}).

\bibitem[{\citenamefont{Hewson}(1993{\natexlab{b}})}]{hewson1}
\bibinfo{author}{\bibfnamefont{A.~C.} \bibnamefont{Hewson}},
  \bibinfo{journal}{Phys. Rev. Lett.} \textbf{\bibinfo{volume}{70}},
  \bibinfo{pages}{4007} (\bibinfo{year}{1993}{\natexlab{b}}).

\bibitem[{\citenamefont{Hewson et~al.}(2006)\citenamefont{Hewson, Bauer, and
  Koller}}]{hewson2}
\bibinfo{author}{\bibfnamefont{A.~C.} \bibnamefont{Hewson}},
  \bibinfo{author}{\bibfnamefont{J.}~\bibnamefont{Bauer}}, \bibnamefont{and}
  \bibinfo{author}{\bibfnamefont{W.}~\bibnamefont{Koller}},
  \bibinfo{journal}{Phys. Rev. B} \textbf{\bibinfo{volume}{73}},
  \bibinfo{pages}{045117} (\bibinfo{year}{2006}).

\bibitem[{\citenamefont{Borda et~al.}(2003)\citenamefont{Borda, Zar\'and,
  Hofstetter, Halperin, and von Delft}}]{borda}
\bibinfo{author}{\bibfnamefont{L.}~\bibnamefont{Borda}},
  \bibinfo{author}{\bibfnamefont{G.}~\bibnamefont{Zar\'and}},
  \bibinfo{author}{\bibfnamefont{W.}~\bibnamefont{Hofstetter}},
  \bibinfo{author}{\bibfnamefont{B.~I.} \bibnamefont{Halperin}},
  \bibnamefont{and} \bibinfo{author}{\bibfnamefont{J.}~\bibnamefont{von
  Delft}}, \bibinfo{journal}{Phys. Rev. Lett.} \textbf{\bibinfo{volume}{90}},
  \bibinfo{pages}{026602} (\bibinfo{year}{2003}).

\bibitem[{\citenamefont{Hofstetter and Zarand}(2004)}]{hofstetter}
\bibinfo{author}{\bibfnamefont{W.}~\bibnamefont{Hofstetter}} \bibnamefont{and}
  \bibinfo{author}{\bibfnamefont{G.}~\bibnamefont{Zarand}},
  \bibinfo{journal}{Phys. Rev. B} \textbf{\bibinfo{volume}{69}},
  \bibinfo{pages}{235301} (\bibinfo{year}{2004}).

\bibitem[{\citenamefont{Oguri and Hewson}(2005)}]{oguri}
\bibinfo{author}{\bibfnamefont{A.}~\bibnamefont{Oguri}} \bibnamefont{and}
  \bibinfo{author}{\bibfnamefont{A.~C.} \bibnamefont{Hewson}},
  \bibinfo{journal}{J. Phys. Soc. Jpn.} \textbf{\bibinfo{volume}{74}},
  \bibinfo{pages}{988} (\bibinfo{year}{2005}).

\bibitem[{\citenamefont{Ng and Lee}(1988)}]{ng-lee}
\bibinfo{author}{\bibfnamefont{T.~K.} \bibnamefont{Ng}} \bibnamefont{and}
  \bibinfo{author}{\bibfnamefont{P.~A.} \bibnamefont{Lee}},
  \bibinfo{journal}{Phys. Rev. Lett.} \textbf{\bibinfo{volume}{61}},
  \bibinfo{pages}{1768} (\bibinfo{year}{1988}).

\bibitem[{\citenamefont{Simon and Affleck}(2001)}]{simon}
\bibinfo{author}{\bibfnamefont{P.}~\bibnamefont{Simon}} \bibnamefont{and}
  \bibinfo{author}{\bibfnamefont{I.}~\bibnamefont{Affleck}},
  \bibinfo{journal}{Phys. Rev.} \textbf{\bibinfo{volume}{64}},
  \bibinfo{pages}{085308} (\bibinfo{year}{2001}).

\bibitem[{\citenamefont{Langreth}(1966)}]{langreth}
\bibinfo{author}{\bibfnamefont{D.~C.} \bibnamefont{Langreth}},
  \bibinfo{journal}{Phys. Rev.} \textbf{\bibinfo{volume}{150}},
  \bibinfo{pages}{516} (\bibinfo{year}{1966}).

\bibitem[{\citenamefont{Haldane}(1978)}]{haldane}
\bibinfo{author}{\bibfnamefont{F.~D.~M.} \bibnamefont{Haldane}},
  \bibinfo{journal}{J. Phys. C: Solid State Phys.}
  \textbf{\bibinfo{volume}{11}}, \bibinfo{pages}{5015} (\bibinfo{year}{1978}).

\bibitem[{\citenamefont{Abrahams et~al.}(1979)\citenamefont{Abrahams, Anderson,
  Licciardello, and Ramakrishnan}}]{abrahams}
\bibinfo{author}{\bibfnamefont{E.}~\bibnamefont{Abrahams}},
  \bibinfo{author}{\bibfnamefont{P.~W.} \bibnamefont{Anderson}},
  \bibinfo{author}{\bibfnamefont{D.~C.} \bibnamefont{Licciardello}},
  \bibnamefont{and} \bibinfo{author}{\bibfnamefont{T.~V.}
  \bibnamefont{Ramakrishnan}}, \bibinfo{journal}{Phys. Rev. Lett.}
  \textbf{\bibinfo{volume}{42}}, \bibinfo{pages}{673} (\bibinfo{year}{1979}).

\bibitem[{\citenamefont{Fleury and Waintal}(2008)}]{fleury-waintal}
\bibinfo{author}{\bibfnamefont{G.}~\bibnamefont{Fleury}} \bibnamefont{and}
  \bibinfo{author}{\bibfnamefont{X.}~\bibnamefont{Waintal}},
  \bibinfo{journal}{Phys. Rev. Lett.} \textbf{\bibinfo{volume}{100}},
  \bibinfo{pages}{076602} (\bibinfo{year}{2008}).

\bibitem[{\citenamefont{Silvestrov and Imry}(2007)}]{silvestrov}
\bibinfo{author}{\bibfnamefont{P.~G.} \bibnamefont{Silvestrov}}
  \bibnamefont{and} \bibinfo{author}{\bibfnamefont{Y.}~\bibnamefont{Imry}},
  \bibinfo{journal}{Phys. Rev. B} \textbf{\bibinfo{volume}{75}},
  \bibinfo{pages}{115335} (\bibinfo{year}{2007}).

\end{thebibliography}
\end{document}